\documentclass[12pt]{article}

\usepackage{amssymb, amsmath, amsthm}
\usepackage{graphicx}
\usepackage{cite,color}

\newcommand{\td}{\text{d}}

\usepackage{hyperref}

\def\be{\begin{equation}}
\def\ee{\end{equation}}
\def\bea{\begin{eqnarray}}
\def\eea{\end{eqnarray}}

\theoremstyle{definition}

\usepackage[left=2cm,right=2cm,top=2cm,bottom=2cm]{geometry}

\title{\bf Distorted static black holes with a bubble }
\author{Matin Tavayef$^d$\footnote{mtavayef@mun.ca}, Shohreh Abdolrahimi $^a$\footnote{sabdolrahimi@cpp.edu} , Ivan Booth$^b$\footnote{ibooth@mun.ca}, and Hari Kunduri$^c$\footnote{kundurih@mcmaster.ca}  \\ \\
\small \sl $^a$Department of Physics and Astronomy, California State Polytechnic University, \\ \small \sl 
Pomona, CA 91768, USA  \\
\small \sl $^b$Department of Mathematics and Statistics, Memorial University of Newfoundland \\ \small \sl 
St. John's, NL,  A1C 5S7, Canada  \\
\small \sl $^c$Department of Mathematics and Statistics \& Department of Physics and Astronomy \\ \small \sl  McMaster University,  Hamilton, ON, L8S 4M1, Canada \\ 
\small \sl $^d$Faculty of Science, Theoretical Physics, Memorial University of Newfoundland \\ \small \sl  St. John's, NL  A1C 5S7, Canada }

\date{}

\begin{document}
\maketitle
\begin{abstract} We construct a family of local static, vacuum five-dimensional solutions with two commuting spatial isometries describing a black hole with a $S^3$ horizon and a 2-cycle `bubble' in the domain of outer communications. The solutions are obtained by adding distortions to an  asymptotically flat seed solution. We show that the conical singularities in the undistorted geometry can be removed by an appropriate choice of the distortion. 
\end{abstract}

\vspace{.5cm}
\section{Introduction} 
A classic result of Lichnerowiscz asserts that the only globally stationary, four-dimensional asymptotically flat solution of the Einstein-Maxwell equations is Minkowski spacetime \cite{Lich} (see also \cite{Gibbons:2011aqq}). Physically, this implies that an isolated gravitating system with positive energy must contain a black hole. A similar result can be proved for electrovacuum \emph{static} solutions in dimensions greater than four \cite{Kunduri:2017htl} (see also \cite{Shiromizu:2012hb}). On the other hand, there are a large number of families of stationary, asymptotically flat solutions that have positive energy but do not contain black holes (for a review, see e.g. \cite{Bena:2007kg}). Such spacetimes may be referred to as gravitational solitons~\cite{Gibbons:2013tqa}. They are characterized by non-trivial 2 cycles, or `bubbles', which are prevented from collapsing by a magnetic flux. Indeed, one can show quite easily using the positive mass theorem and Stokes' theorem that in the pure vacuum case, with vanishing Maxwell fields, such solutions cannot exist.  Moreover, it can be proved that the assumption of trivial topology on Cauchy surfaces is sufficient to rule out the existence of solitons~\cite{Kunduri:2013vka}. 

Consider the existence of asymptotically flat stationary spacetimes containing a black hole which has such non-trivial topology in its domain of outer communication. This can be interpreted as an equilibrium configuration of a black hole and a soliton. Such solutions would produce a continuous failure of black hole uniqueness, even for fixed horizon topology, as one could not distinguish between two black holes with the same conserved charges computed at asymptotic infinity. Explicit examples of such configurations have indeed been constructed, confirming this continuous violation of uniqueness for topologically spherical black holes\cite{Horowitz:2017fyg, Breunholder:2017ubu}. The non-trivial topology leads to new terms in the first law of black hole mechanics~\cite{Kunduri:2013vka}. The known examples are all supersymmetric, where there is a great deal known about the local form of solutions~\cite{Gauntlett:2002nw}. They have non-zero charge and angular momenta.  Nonetheless, it is natural to expect that non-supersymmetric, and possibly even pure vacuum, black holes spacetimes with non trivial topology in the exterior region should exist. Indeed, only recently, an explicit example of a family of non-supersymmetric black hole solutions of this type in minimal supergravity was constructed \cite{Suzuki:2023nqf}. The static black hole uniqueness theorem states that all (electro-)vacuum, asymptotically flat black holes are members of the (Reissner-Nordstrom) Schwarzschild family \cite{Gibbons:2002av}.  Hence any stationary, asymptotically flat vacuum black hole spacetime with a bubble would have to be non-static. Explicitly constructing smooth solutions of this type is well known to be quite difficult. 

To gain some insight on this problem, one could consider relaxing one or more of the conditions, such as asymptotic flatness or non-staticity. It is fairly straightforward to construct explicit solutions describing a black hole-bubble configuration within the Weyl class of static, biaxisymmetric solutions (those admitting a $\mathbb{R} \times U(1) \times U(1)$ isometry group with mutually orthogonal generators)~\cite{GeneralizedWeyl}. The resulting local metrics can be chosen to be asymptotically flat, but they suffer from conical singularities associated to the fixed point sets (`axes') of the rotational Killing vector fields (see Section 2). The conical singularities can be chosen to lie either in the interior of the spacetime or at infinity. As shown by Tomlinson~\cite{Tomlinson:2021wsp}, Harrison-type transforms can be used to produce new solutions of the Einstein-Maxwell equations from this seed solution. The new solutions will have additional parameters which could in principle be used to eliminate the conical singularities, although the regular solutions will no longer be asymptotically flat. Such a strategy was carried out in detail in~\cite{Tomlinson:2021wsp}, although it was found the parameters could not be appropriately chosen to achieve regularity. In this article, we will focus on a different approach to achieve regularity: relaxing the condition of asymptotic flatness. As we explain below, this can be thought of as including external matter in the asymptotic region, which distorts the geometry near the black hole. 

Most of our understanding of stationary black holes, their properties and structures come from studying black holes with maximal symmetry, such as Schwarzschild, Kerr, or Reissner-Nordstrom-like black hole solutions. In analogy with the theory of electromagnetism, the study of such black holes is analogous to the study of a single charge in an empty space where no interaction is present. We may wonder to what extent the properties of such black holes are a reliable guide to those of more general solutions. Are there new unexpected properties when we study the interactions of black holes with surrounding matter and sources? We can think of large black holes as long wavelength IR backgrounds where Einstein’s classical equations are reliable. However, the large black holes have so much information even though they are IR objects. From this perspective, as stated by Agmon et. al. \cite{Agmon} ``black holes are the star of the show in many aspects of quantum gravity." 

Lately, many of the Swampland conditions are also motivated by black hole physics, although some others just have string theory backing \cite{Agmon}. These new developments, although not the only motivation, enhance the need for study of more general class of black hole solutions. For example, it has been shown that some of the features of black holes are not as universal as we might think. In particular, it was demonstrated that, in the case of a distorted five-dimensional Myers-Perry black hole, the ratio of the horizon angular momentum and the mass $J^2/M^3$ is unbounded, and can grow arbitrarily large \cite{Abdolrahimi2014}. Similarly, for a distorted Kerr black hole, the solution is regular outside the horizon even though the spin parameter can satisfy $J^2/M^4>1$ \cite{Abdolrahimi2015}. There have also been efforts in investigating various aspects of more general black holes. 
For example, a heuristic argument for the universal area relation
of a four-dimensional adiabatically distorted Kerr-Newman black hole has been proposed \cite{Ansorg2008,Cvetic,Page2015}. 

Recently, it was illustrated that some deformed black holes with less symmetry are more stable against more general class of perturbations \cite{Horowitz2022}. Thus, studying distorted black holes is crucial as at times their properties diverge from the universal characteristics observed in symmetric black hole solutions, challenging our understanding and highlighting the need for a more comprehensive exploration of black hole physics to capture the diverse phenomena arising from interactions with surrounding matter and sources. Although the ideal situation would be to analyze dynamical black holes, there exists a well-known technique for constructing distorted static black hole solutions. This method relies on the fact that for the Weyl class of metrics, the Einstein equations reduce to solving for axisymmetric solutions of Laplace's equation on $\mathbb{R}^3$ \cite{Weyl}. The resulting linearity can be used to `deform' a given Weyl solution by adding harmonic functions. In higher dimensions, using the generalized Weyl form \cite{GeneralizedWeyl}, we can construct distorted higher dimensional black holes.

In Newtonian gravity, multipole expansions are commonly used to expand the potential 
associated with a particular mass distribution. Around a central point  
such a series
can be written in terms of both positive and negative powers of a radial function $r$ (the origin is $r=0$). However it is 
typically applied to just one of two situations: 1) if the mass is localized to a small region then the expansion is in terms of \emph{exterior multipole moments}: negative powers so that the potential
goes to zero at infinity or 2) if the mass is in a shell far from the origin then 
the expansion is in terms of \emph{interior multipole moments}: positive powers so that potential diverges at infinity. 
% Multipole expansions are used frequently in the study of electromagnetism and the gravitational field in Newtonian gravity. A multipole expansion provides an exact description of the potential under the following two conditions: (1) if the sources (e.g. charges/masses) are localized close to the origin and the point at which the potential is observed is far from the origin; or the reverse, i.e., (2) if the sources are located far from the origin and the potential is observed close to the origin. In the first case, the coefficients of the series expansion are called exterior multipole moments or simply multipole moments whereas, in the second case, they are called interior multipole moments. 
It is in this context that we consider a local black hole solution. Namely, as pointed out by Geroch and Hartle \cite{GerochH}, the procedure can be interpreted physically as adding new external `sources' from an isolated self-gravitating system that distort the interior geometry\footnote{{As discussed in \cite{GerochH}, there is no guarantee that these `sources' will not violate the energy conditions.
The allowed values of the multipole moments will be constrained if one wishes to only consider distortions that can be imposed by non-energy-condition violating matter. Exact constraints must be determined
on a case-by-case basis. 
%Further conditions may need to be imposed on the allowed values of multiple moments.
}}. In this case, in analogy with the multiple expansion in electromagnetism we don't include exterior sources in the solution. Thus, the solution is valid only in the interior, local region near the black hole and suffers from lack of asymptotic flatness. This solution-generating technique has been used to produce and study many new solutions, generally known as distorted black holes or black objects, such as distorted Schwarzschild, Kerr or Reissner Nordstr\"om solutions, distorted Myers-perry black hole, and static and charged black rings, etc. \cite{dis1,dis2,dis3,dis4,dis5, Fairhurst, Israel1776, Tomimatsu, Breton1997, Breton1998, Ansorg2008, Abdolrahimi2009, Ansorg2009, Ansorg2009-2, Abdolrahimi2010, Abdolrahimi2013, Abdolrahimi2014, Abdolrahimi2015, Abdolrahimi2015B,Shoom2015A, Shoom2015, Basovnik2016, Grover2018, Abdolrahimi2020,Deligianni2020, Marco2021,Abdolrahimi2015C}. In this paper we construct a distorted local static black hole-bubble solution that is smooth everywhere on and outside the horizon and is free of conical singularities. 
\section{The black hole-bubble solution}
  We first describe a vacuum solution describing a black hole with a non-collapsing $S^2$ `bubble' in the domain of outer communication. The geometry is asymptotically flat, but the metric suffers from two conical singularities along two axes of symmetry. We briefly describe the construction of such metrics below. 
\subsection{Weyl solutions} The general solution of the $D$-dimensional vacuum Einstein equations admitting $D-2$ orthogonal commuting (non-null) Killing vector fields is given in terms of $D-3$ independent axisymmetric solutions of Laplace's equation in Euclidean $\mathbb{R}^3$.  Such solutions are known as `Weyl' solutions based on the classic original work in $D=4$.  We will focus on the $D=5$ case. We refer to the reader to the review \cite{Weyl, GeneralizedWeyl} for details. 

A five dimensional Weyl metric can be locally expressed in the form
\begin{equation}
	\td s^2=-e^{2U_{0}}\td t^2+e^{2\nu}(\td r^{2} + \td z^{2})+e^{2U_{1}}\td\psi^{2}+e^{2U_{2}}\td\phi^{2}\label{GenWeylMetric}
\end{equation} Here $t \in \mathbb{Z}$ is a timelike coordinate, $r > 0$, $z \in \mathbb{R}$, and $\psi, \phi$ will be chosen to each be identified with period $2\pi$ in order to obtain an asymptotically flat geometry and /or a regular solution.  The Killing vector fields $\partial_t, \partial_\psi, \partial_\phi$ generates the commuting isometry group $\mathbb{R} \times U(1) \times U(1)$.  The Ricci flat equations imply that the metric functions $U_i = U_{i}(r,z)$, $i=0,1,2$ are each axisymmetric solutions of the Laplace equation in $\mathbb{R}^3$, which in cylindrical coordinates $(r,z,\theta)$ reads
\begin{equation}\label{Laplace}
	\dfrac{\partial^2{U_{i}}}{\partial r^2}+\frac{1}{r}\dfrac{\partial{U_{i}}}{\partial r}+\dfrac{\partial^2{U_{i}}}{\partial z^2}=0 \; . 
\end{equation} Here $(r,z)$ is identified with the orbit space of the original spacetime by the isometry group and $\theta \sim \theta + 2\pi$ is an auxiliary coordinate.  Equivalently,  $ \nabla^{2} U_{i}=0$ where $\nabla^2$ is the Laplacian associated to flat $\mathbb{R}^3$ with metric
\begin{equation}
	\td s^2=\td r^2+r^{2}\td \theta^{2} +\td z^2 \; .
\end{equation} Note that, the three $ U_{i} $ functions are not all independent, but satisfy the constraint
\begin{equation}\label{grsr}
	\sum_{i}U_{i}=\log r+ c \; .
\end{equation}
We can adjust the constant term $c$ freely by rescaling the coordinates $ x^{i} $.  It's important to note that  $ \log r $ represents a solution to Laplace's equation, which is the Newtonian potential sourced by a one-dimensional rod with an infinite length along the $z-$axis,  having a uniform mass density of 1/2 . Once two harmonic functions $U_i$ are selected, the Weyl metric is locally fully determined. The remaining function $\nu = \nu(r,z)$ satisfies the first order equations 
\begin{align}\label{nur}
	\partial_{r} \nu &=-\frac{1}{2r}+\frac{r}{2}\sum_{i=0}^{2}\left((\partial_{r}U_{i})^{2}-(\partial_{z}U_{i})^{2}\right) \\
	\partial_{z} \nu &=r\sum_{i=0}^{2}\left((\partial_{r}U_{i})(\partial_{z}U_{i})\right) \; . \label{nuz}
	\end{align}  The integrability condition for \eqref{nur} and \eqref{nuz} reduces to \eqref{Laplace} and so $\nu$ is determined up to quadrature. 

Let $\hat{g}$ denote the restriction of the metric to the Killing vector fields $(\partial_t, \partial_\psi, \partial_\phi)$. Observe that
\begin{equation}
\det \hat{g} = - e^{2U_0 + 2U_1 + 2U_2} = (e^{2c}) r^2.
\end{equation} The set $r =0$, namely the $z-$ axis,  can be shown to decompose into intervals characterized by whether the timelike field $\partial_t$ is null or a linaer combination of the spatial Killing fields $(\partial_\psi, \partial_\phi)$ degenerate.  The resulting submanifolds in the spacetime correspond to an event horizon or an axis of symmetry of either of the Killing fields. The points in the orbit space where two spacelike axis intervals (referred to as `rods') intersect are called `corners'. At these points the $U(1)^2$ torus action degenerates. Potential conical singularities arise on these axes and the parameters of the solutions must be chosen to achieve a globally smooth metric. This latter step cannot always be performed. In particular, this is the case for the asymptotically flat black hole-bubble solution which we now describe. We refer the reader to \cite{GeneralizedWeyl} where a large class of examples are discussed in detail. 
\subsection{Unbalanced black hole-bubble solution}
As described above, an event horizon in the full spacetime correspond to a (finite) timelike rod $z_a < z < z_b$ on which $e^{2U_0} =0$. A finite spacelike axis rod with corners as endpoints,  where one of $e^{2U_i}$, $i = 1,2$ vanishes, corresponds to a topological $S^2$ (the $S^1$ corresponding to the generator associated to $U_i$ degenerates) with poles at the corner points. Such $S^2$ surface lie in the region connected to the asymptotic region $r^2 + z^2 \to \infty$ and represent `bubbles'  in the black hole exterior.  Another possibility for a finite spacelike rod is that one in which its endpoints intersects the event horizon rod; such a rod describes a topological disc (hemisphere) in the full spacetime. 
Furthermore, for a five-dimensional asymptotically flat metric, there must be two semi-infinite axes $(-\infty, z_1)$ and $(z_2, \infty)$ on which $\partial_\psi$ and $\partial_\phi$ degenerate respectively; these correspond in the spacetime to the poles of the asymptotic $S^3$ boundary at spatial infinity. For a given spacetime in the Weyl class, the associated collection of horizon and axis rods is referred to as its rod structure. 

One can construct a simple example of an asymptotically flat black hole with a $S^3$ horizon and a bubble in the exterior in the region within the Weyl class \cite{Rose} as follows. Let $a,b,c$ be positive constants satisfying $0 < a < b < c$ and define
\begin{equation}\label{muk}
	\mu_{k}=\sqrt{r^{2}+(z-k)^{2}}-(z-k)    
\end{equation} 
Observe that $\log \mu_k$ is an axisymmetric harmonic function on $\mathbb{R}^3 / \mathcal{A}$ where $\mathcal{A}$ represents the region on the $z-$ axis where $\mu_k =0$, i.e. $\{(r,z) |r=0, z > k\}$; it is the Newtonian potential for a semi-infinite rod $z>k$ of mass density $1/2$.  It is a simple to construct a Weyl solution by superimposing these potentials and using the linearity of the Laplace equation.  The black hole-bubble solution we will consider has harmonic functions $U_i$ defined by
\begin{equation}\label{Ui}
	e^{2U_{0}}=\dfrac{\mu_{0}}{\mu_{a}}, \quad e^{2U_{1}}=\dfrac{r^{2}\mu_{b}}{\mu_{0}\mu_{c}}, \quad e^{2U_{2}}=\dfrac{\mu_{a}\mu_{c}}{\mu_{b}}.
\end{equation} From this choice we may read off the rod structure (see Figure 1):
\begin{enumerate}
\item $I_1$: a semi-infinte rod $-\infty < z < 0$ corresponding to the asymptotic symmetry axis of $\partial_\psi$
\item $H$: a finite timelike $0 < z < a$ corresponding to an event horizon with spatial cross section $S^3$.
\item $I_D$: a finite spacelike rod $a < z < b$ on which $\partial_\phi$ degenerates. This represents a disc. 
\item $I_B$: a finite spacelike rod $b < z < c$ on which $\partial_\psi$ degenerates. This represents a bubble as explained above. 
\item $I_2$ a semi-infinte rod $c < z < \infty$ corresponding to the asymptotioc symmetry axis on which $\partial_\phi$ degenerates. 
\end{enumerate}
\begin{figure}[t]
	\centering
	\includegraphics[width=0.7\textwidth]{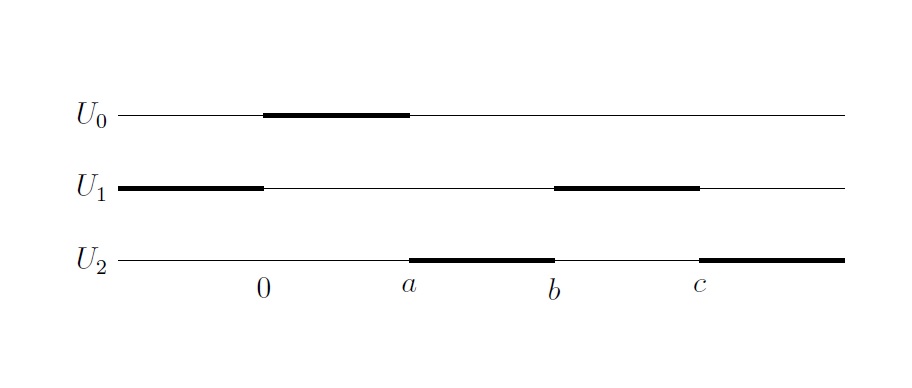}
	\caption{The rod structure for a black hole with a bubble \cite{Rose}. Here the finite rod along (0 $<$ z$ < $a) corresponds to a static black hole horizon with spatial cross-section $S^3$.}
	\label{fig4}
\end{figure} This solution represents an asymptotically flat black hole with horizon cross section $S^3$ with a non-trivial 2-cycle (bubble) in the domain of outer communications, i.e. the exterior region to the black hole from which light signals can escape to the asymptotic region\cite{Rose, Tomlinson:2021wsp}.  

To write the metric in the form  (\ref{GenWeylMetric}) requires an explicit expression for the function $\nu$.  We follow the approach of Iguchi, Izumi, and Mishima \cite{IM}. Define the Euclidean distance from a point in the orbit space from $(0,z)$: 
\begin{equation}
	R_{c}=\sqrt{r^{2}+(z-c)^2}
\end{equation} and 
\begin{equation}
	\bar{\mu_{c}}=R_{c}+(z-c)
\end{equation} along with $\bar{U_{c}}=\frac{1}{2}\log(R_{c}+(z-c))$ so that $\bar\mu_c = e^{2\bar{U}_c}$.
\begin{equation}
	\gamma_{cd}=\frac{1}{2}\bar{U_{c}}+\frac{1}{2}\bar{U_{d}}-\frac{1}{4}\log Y_{cd}, \qquad Y_{cd}=R_{c}R_{d}+(z-c)(z-d)+r^{2}
\end{equation} Then the $ \gamma_{cd} $ satisfy the first order equations
\begin{equation}\label{dgammar}
	\partial_{r}\gamma_{cd}=r\left(\partial_{r}\bar{U_{c}}\partial_{r}\bar{U_{d}}-\partial_{z}\bar{U_{c}}\partial_{z}\bar{U_{d}}\right),
\end{equation}
\begin{equation}\label{dgammaz}
	\partial_{zr}\gamma_{cd}=r\left(\partial_{r}\bar{U_{c}}\partial_{z}\bar{U_{d}}+\partial_{z}\bar{U_{c}}\partial_{r}\bar{U_{d}}\right).
\end{equation} For the black hole-bubble solution, we can rewrite the $U_i$ in terms of the $\bar{U}_i$:
\begin{equation}\label{Ui2}
	U_{0}=\bar{U_{a}}-\bar{U_{0}},\quad  U_{1}=\bar{U_{0}}+\bar{U_{c}}-\bar{U_{b}},\quad
	U_{2}=\log r-\bar{U_{a}}-\bar{U_{c}}+\bar{U_{b}}.
\end{equation} This leads to the relations
\begin{equation}\label{expUi2}
	e^{2U_{0}}=\dfrac{\bar{\mu}_{a}}{\bar{\mu}_{0}}, \quad e^{2U_{1}}=\dfrac{	\bar{\mu}_{0}	\bar{\mu}_{c}}{\bar{\mu}_{b}}, \quad e^{2U_{2}}=\dfrac{r^{2}\bar{\mu}_{b}}{\bar{\mu}_{c}\bar{\mu}_{a}}.
\end{equation} It remains a tedious process to rewrite $\td \nu$ in terms of the $\bar{U}_i$. Doing so and comparing to \eqref{dgammar}, \eqref{dgammaz} one finds \cite{Rose}
\begin{equation}\label{nuAF}
	\nu(r,z)=\bar{U_{b}}-\bar{U_{a}}-\bar{U_{c}}+\gamma_{00}+\gamma_{aa}+\gamma_{bb}+\gamma_{cc}-\gamma_{0a}+\gamma_{0c}-\gamma_{0b}-2\gamma_{bc}+\gamma_{ac}-\gamma_{ab}+C
\end{equation} where $C$ is an integration constant which is fixed by requiring asymptotic flatness. Consider the flat Euclidean metric on $\mathbb{R}^4$ in spherical coordinates $(R^{\prime},\theta,\phi_1,\phi_2)$ 
\begin{equation}
    \delta_4 = \td R^{\prime 2} + R^{\prime2} (\td \theta^2 + \sin^2\theta \td \phi_1^2 + \cos^2\theta \td \phi_2^2).
\end{equation} with $R^{\prime}>0$, $\theta \in (0,\pi/2)$ and $\phi_i$, $i=1,2$ are angles with period $2\pi$. We can rewrite this in the Weyl form by identifying $\phi_1 = \psi,  \phi_2 = \phi$ and using the transformation
\begin{equation}
R^{\prime}= \left[ 4(r^2 + z^2) \right]^{1/4}, \qquad \tan 2\theta = \frac{r}{z} 
\end{equation} which has inverse
\begin{equation}
r=\dfrac{R^{\prime2}\sin2\theta}{2} \qquad z=\dfrac{R^{\prime2}\cos 2\theta}{2}
\end{equation} one finds that the Euclidean metric on $\mathbb{R}^4$ is
\begin{equation}\begin{aligned}
\delta_4= \frac{\td r^2 + \td z^2}{2\sqrt{r^2 + z^2}} + (\sqrt{r^2 + z^2} + z) \td \psi^2 +  (\sqrt{r^2 + z^2} - z) \td \phi^2.
\end{aligned}
\end{equation}  Thus for a Weyl metric to be asymptotically flat, we require that as $R^{\prime} \to \infty$, $e^{2\nu} = \sqrt{2}e^{2C} R^{\prime-2} + O(R^{\prime-4})$.  Comparing this to \eqref{nuAF} fixes
\begin{equation}\label{C}
C = -\frac{1}{4} \log 2.
\end{equation} 
To summarize, the local form of the metric describing an asymptotically flat black hole with a bubble in the exterior is given in the Weyl form \eqref{GenWeylMetric} with metric functions \eqref{Ui} and conformal factor $\nu$ given in \eqref{nuAF} and $C$ fixed by \eqref{C}. 

We now turn to global properties of the metric. The functions \eqref{Ui} are analytic everywhere away from the singular sets of the $U_i$ (being compositions of the exponential function and harmonic functions) and hence the metric is analytic as well away from the singular set defined by $r=0$.  Each $U_i$ becomes singular on some interval $(z_{i-1}, z_i)$ of the $z-$axis, corresponding to the horizon or axes. These intervals meet at rod points $z_i$ where two of the $e^{2U_i}$ simultaneously vanish. Regularity of the metric at these points is guaranteed provided certain admissibility conditions are met (namely, that the determinant of the matrix formed by the associated rod vectors is $\pm 1$) \cite{Harmark}. This condition is automatically satisfied for Weyl solutions as the rod vectors are orthogonal. 

Potential singularities may arise on the interior of each interval upon which a given $U_i$ behaves like $\log r$ as $r \to 0$. The singular set of $U_0$ corresponds to the event horizon where $\partial_t$ becomes null. Our choice of $U_0$ is sufficient to guarantee that the metric will be smooth across the horizon \cite{Harmark}. The other types of singularity that can arise are conical singularities associated to fixed point sets of the rotational Killing fields. Consider an axis rod (a codimension 2 surface in spacetime), upon which a spatial Killing field $K$ vanishes. Let $\phi$ be an adapted coordinate so that $K = \partial_\phi$. A smooth degeneration of $K$ (removal of conical singularities) requires that $\phi \sim \phi + \Delta \phi$ where
\begin{equation}\label{conical}
\Delta \phi = 2\pi \lim_{r \to 0} \sqrt{ \frac{r^2 e^{2\nu}}{|K|^2}}
\end{equation} For a solution with multiple axes for a given $K = \partial_\psi, \partial_\phi$, it is not generally possible to choose $\Delta \phi, \Delta \psi$ so that all conical singularities are removed. 

Along the two semi-infinite rods, it is easily checked that
\begin{equation}
\sqrt{\lim\limits_{r\to 0}	\dfrac{r^{2} e^{2\nu}}{e^{2U_{2}}}}=1, \qquad z > c \qquad \sqrt{\lim\limits_{r\to 0}	\dfrac{r^{2} e^{2\nu}}{e^{2U_{1}}}}=1, \quad z < 0,
\end{equation} which fixes the periodicities $\Delta \psi = \Delta \phi = 2\pi$. This is necessary due to the requirement that the spacetime is asymptotically flat, and hence there is an asymptotic round $S^3$ boundary at spatial infinity. 

 For a black hole with a bubble, there are also two finite axis rods as shown in the figure $ (\ref{fig4}) $. Each finite axis rod in this solution is associated to a conical singularity. First, we examine the case of $I_D$, the disc topology region corresponding to $r=0, a < z < b$ along which $\partial_\phi$ degenerates.  A calculation shows
\begin{equation}\label{nonAFphi}
\sqrt{\lim\limits_{r\to 0}	\dfrac{r^{2} e^{2\nu}}{e^{2U_{2}}}}=\sqrt{\dfrac{b(b-a)}{c(c-a)}}, \quad a<z<b<c
\end{equation} Regularity requires that this be equal to unity, implying $ b+c=a $. However since by assumption $ 0<a<b<c $, this cannot be achieved.  Next along the bubble rod $I_B$ where $r=0,  b < z < c$ along which $\partial_\psi$ degenerates, we find
\begin{equation}\label{nonAFpsi}
\sqrt{\lim\limits_{r\to 0}\dfrac{r^{2} e^{2\nu}}{e^{2U_{1}}}}=\sqrt{\dfrac{(c-b)^2}{c(c-a)}}, \quad b<z<c
\end{equation}  However it is clear that
\begin{equation}
	\sqrt{\dfrac{(c-b)^2}{c(c-a)}}<1
\end{equation} Therefore there is no way to choose the parameters  $(a,b,c)$ to remove the conical singularities while keeping the rod structure fixed, i.e. $0 < a < b < c$.   One could choose the periodicities of the angles so that the conical singularities are removed on the disc $a < z < b$ and on the bubble $b < z < c$ although the resulting solution would have conical singularities along the semi-infinite rods that extend to spatial infinity. 

Before we turn to constructing the distorted solution, we give expressions for some geometric invariants of physical interest.  From the expansion 
\begin{equation}
	g_{tt}(R',\theta)=-1+\dfrac{2(a)}{R'^{2}}+O(R'^{-4})
\end{equation} one can read off the Komar mass
\begin{equation}
	m=\dfrac{3\pi a}{4},
\end{equation} which must agree with the ADM mass for stationary, asymptotically flat solutions. Note that the mass is independent of the parameters $(b,c)$.  The surface gravity of the event horizon is 
\begin{equation}
\kappa = \lim_{r\to 0} \sqrt{\frac{e^{2U_0}}{r^2 e^{2\nu}}} = \sqrt{\frac{c}{2ab}}
\end{equation} where $0 < z < a$ on the horizon rod.  Spatial cross sections of the horizon have topology $S^3$ with an inhomogeneous metric.  As discussed above, here will be a conical singularity at one pole (at the pole $z=a$ where $\partial_\phi$ vanishes). The area of the horizon is
\begin{equation}
\text{A}_H = \int _H \td \mu =4\pi^2  \int_0^a e^{\nu + U_1 +  U_2} \td z = 4\pi^2 a \sqrt{\frac{2ab}{c}}.
\end{equation}  Observe that the Smarr relation
\begin{equation}
m = \frac{3 \kappa A_H}{16\pi} = \frac{3 \pi a}{4} . 
\end{equation} holds.  The Komar mass on the horizon, 
\begin{equation}\label{Komar}
m_H =  - \frac{3}{32\pi} \int_H \star \td k = \frac{3\pi}{4} \int_0^a \partial_r U_0 e^{U_0 +  U_1 + U_2} \; dz = \frac{3\pi}{4} \cdot a
\end{equation} which equals the mass computed at infinity, as it must by Stokes' theorem. The area of the $S^2$ bubble is
\begin{equation}
\text{A}_B = 2\pi \int_b^c e^{\nu + U_2} \; \td z = 2\pi (c-b) \left[ 1 - \sqrt{\frac{b(b-a)}{c(c-a)}} + \frac{a}{\sqrt{c(c-a)}} \log \left[ \frac{\sqrt{b} - \sqrt{b-a}}{\sqrt{c} - \sqrt{c-a}} \right] \right],
\end{equation}  and the area of the disc (hemisphere) is
\begin{equation}
\text{A}_D = 2\pi \int_a^b e^{\nu + U_1} \; \td z = \frac{2\pi}{\sqrt{c(c-a)}} \left[b -a + a \sqrt{b(b-a)} \sinh^{-1} \left(\frac{b}{a} - 1\right) \right]. 
\end{equation} 
\subsection{The event horizon} The Weyl coordinate system covers only the region of spacetime outside any event horizons. Note that $g_{tt} = -e^{2U_0} < 0$ and it vanishes in the limit $r \to 0$ and $z \in (0,a)$. We can pass to a natural radial type coordinate by introducing coordinates $(\rho,x)$
\begin{equation}
    r^2 = \left(\rho^2 - \frac{a^2}{4} \right)(1-x^2), \qquad z = \frac{a}{2} + \rho x
\end{equation} In this coordinate chart, the horizon corresponds to $\rho = a/2$ and $|x| \leq 1$.  Under this transformation, 
\begin{equation}
    e^{2\nu} (\td r^2 + \td z^2) = e^{2\nu} \left(\rho^2 - \frac{a^2 x^2}{4}\right) \left[\frac{\td \rho^2}{\rho^2 - \frac{a^2}{4}} + \frac{\td x^2}{1 - x^2} \right]. \end{equation} The metric degenerates at the horizon endpoints $\rho = a/2, x =\pm 1$.  In the region near the horizon, the $(t,r)$ part of the metric can be expressed in terms of $(\rho,x)$ coordinate chart as 
    \begin{equation}
        \td s^2_2 = -\frac{\rho - \frac{a}{2}}{a} \left[1 + O(\rho - a/2)^2\right] \td t^2 + \frac{\td\rho^2}{\rho - \frac{a}{2}} \left[\frac{b}{2c} + O(\rho - a/2)^2\right]
    \end{equation} We can then introduce coordinates that are regular on the event horizon by 
    \begin{equation}
        v = a^{-1/2} \left[t + \sqrt{\frac{ab}{2c}} \log \left(\rho - a/2\right)\right]
    \end{equation} in which case the metric takes the form
    \begin{equation}
        \td s^2_2 = -\left(\rho - \frac{a}{2}\right) \td v^2 + 2\sqrt{\frac{b}{2c}} \td v \td \rho + \ldots 
    \end{equation} where the $\ldots$ denote terms that are smooth as $\rho \to a/2$. We thus conclude that we can analytically continue the metric through the (non-degenerate) event horizon into the black hole interior region.  
    
    The geometry of the horizon is given by 
\begin{equation}
   \td s^2_h =  \frac{a b}{2zc(a - z)} \td z^2 + \frac{2 (b-z) z}{c-z} \td \psi^2 + \frac{2(a-z)(c-z)}{b-z} \td \phi^2 .
\end{equation} This can written in a simpler form by writing $x = \cos 2\theta$ with $0 < \theta < \pi/2$, giving 
\begin{equation}
    \td s^2_h = 2a \left[ \frac{b}{c} \td \theta^2 + \frac{(b -a \cos^2\theta)\cos^2\theta}{c -a \cos^2\theta} \td \psi^2 + \frac{ (c -a \cos^2\theta) \sin^2\theta}{b - a\cos^2\theta} \td \phi^2\right].
\end{equation} This can be extended to an inhomogeneous metric on $S^3$, although it necessarily has a conical singularity at one the two poles at $\theta =0$ or $\theta = \pi/2$.

\section{Distorted black hole-bubble solution}
\subsection{Distorting Weyl solutions} For the Weyl class of solutions, the construction of a full vacuum equations reduce to selecting two axisymmetric harmonic functions in $\mathbb{R}^3$. The Laplace equation is obviously linear, and so it is relatively simple to superimpose solutions. In particular, one can `add' higher order axisymmetric harmonics to an existing solution in such a way that the underlying rod structure (but not the geometry) is unaffected. 

Although one can formally consider the solution even at the infinity. Such consideration would lack proper physical interpretation. As we discussed in the introduction, we have two choices either to analyze the spacetime in the interior region of the sources or the exterior region of the sources. This is in analogy with the interior or exterior multiple expansions in the electromagnetism. We choose the interior multiple moments. Thus, our solution is valid only in the interior region, i.e. a local distorted black hole by presence of external sources. In such analysis, the solution is not asymptotically flat. In other words, between the black hole horizon and the asymptotic infinity the sources are located. However, the vacuum Einstein equations we use here to construct the distorted black hole solution can't account for the sources and thus are valid only in the interior region of sources far away from the sources. As we will see, this can help to remove conical singularities in the black hole-bubble system, at the cost of asymptotic flatness. 

Consider a fixed Weyl solution $(\tilde{U}_i, \tilde\nu)$, referred to as the background solution.  We consider a solution $(U_i, \nu)$ defined by
\begin{equation}
	U_{i}=\tilde{U_{i}}+\widehat{U_{i}} , \qquad \nu=\tilde{\nu}+\widehat{\nu}
\end{equation} where $(\widehat{U}_i, \widehat\nu)$ are understood to represent the deformation.  We have
\begin{equation}\label{deformconstraint}
	\tilde{U_{0}}+	\tilde{U_{1}}+	\tilde{U_{2}}=\log r,\qquad	\widehat{U_{0}}+	\widehat{U_{1}}+	\widehat{U_{2}}=0\quad.
\end{equation} It is convenient to reparameterize the background and deformation fields in terms of unconstrained fields as follows: 
\begin{equation}\label{tildeUW}
		\tilde{U_{0}}=	\tilde{U}+	\tilde{W}+\log r, 	 \quad  \tilde{U_{1}}=-\tilde{W}, \quad \tilde{U_{2}}=-\tilde{U}
\end{equation}
\begin{equation}\label{tildenu}
	\tilde{\nu}=\tilde{V}+\tilde{U}+\tilde{W}
\end{equation}
\begin{equation}
	\widehat{U_{0}}=\widehat{U}+\widehat{W},\quad \widehat{U_{1}}=-\widehat{W}, \quad\widehat{U_{2}}=-\widehat{U}~,
\end{equation}
\begin{equation}\label{nuVUW}
	\widehat{\nu}=\widehat{V}+\widehat{U}+\widehat{W}\quad.
\end{equation} The Weyl metric is then expressed as 
\begin{equation}\label{GWS distorted}
	ds^{2}=e^{2(\tilde{U}+\tilde{W}+\widehat{U}+\widehat{W})}\left(-r^{2}\td t^{2}+e^{2(\tilde{V}+\widehat{V})}(\td z^{2}+\td r^{2})\right)+e^{-2(\tilde{W}+\widehat{W})}\td\psi^{2}+e^{-2(\tilde{U}+\widehat{U})}\td\phi^{2}~.
\end{equation} The background fields $\tilde{U}, \tilde{W}$ are both harmonic whereas  $\tilde{V}$ is determined from 
\begin{align}
\tilde{V}_{,r}&=r\left(\tilde{U}_{,r}^{2}+\tilde{W}_{,r}^{2}+\tilde{U}_{,r}\tilde{W}_{,r}-\tilde{U}_{,z}^{2}-\tilde{W}_{,z}^{2}-\tilde{U}_{,z}\tilde{W}_{,z}\right) \\
	\tilde{V}_{,z}&=r\left(2\tilde{U}_{,r}\tilde{U}_{,z}+2\tilde{W}_{,r}\tilde{W}_{,z}+\tilde{U}_{,r}\tilde{W}_{,z}+\tilde{U}_{,z}\tilde{W}_{,r}\right)\quad.
\end{align} In addition, the deformation fields $ \widehat{U} $ and $ \widehat{W} $ are harmonic, and $ \widehat{V} $ can be obtained by using the following integrable equations
\begin{equation}\label{VzVrho}
	\begin{aligned}
		\widehat{V}_{, r} &=r\left(\widehat{U}_{, r}^{2}+\widehat{W}_{, r}^{2}+\widehat{U}_{,r} \widehat{W}_{, r}-\widehat{U}_{, z}^{2}-\widehat{W}_{, z}^{2}-\widehat{U}_{, z} \widehat{W}_{, z}\right.\\
		&+\widetilde{U}_{, r} \widehat{W}_{, r}+\widetilde{W}_{,r} \widehat{U}_{,r}-\widetilde{U}_{, z} \widehat{W}_{, z}-\widetilde{W}_{, z} \widehat{U}_{, z} \\
		&\left.+2\left[\widetilde{U}_{,r} \widehat{U}_{,r}+\widetilde{W}_{,r} \widehat{W}_{,r}-\widetilde{U}_{, z} \widehat{U}_{, z}-\widetilde{W}_{, z} \widehat{W}_{, z}\right]\right) \\
		\widehat{V}_{, z} &=r\left(2 \widehat{U}_{, r} \widehat{U}_{, z}+2 \widehat{W}_{,r} \widehat{W}_{, z}+\widehat{U}_{,r} \widehat{W}_{, z}+\widehat{U}_{, z} \widehat{W}_{, r} \right. \\
		& +\widetilde{U}_{,r} \widehat{W}_{, z}+\widetilde{U}_{, z} \widehat{W}_{,r}+\widetilde{W}_{,r} \widehat{U}_{, z}+\widetilde{W}_{, z} \widehat{U}_{,r} \\
		&\left.+2\left[\widetilde{U}_{, r} \widehat{U}_{, z}+\widetilde{U}_{, z} \widehat{U}_{, r}+\widetilde{W}_{,r} \widehat{W}_{, z}+\widetilde{W}_{, z} \widehat{W}_{,r}\right]\right) \quad.
	\end{aligned}
\end{equation} Choosing the black hole-bubble solution as the background, we read off
\begin{equation}\label{Ut}
\tilde{U}=-\tilde{U_{2}}=\frac{1}{2}\log\left(\dfrac{\bar{\mu_{c}}\bar{\mu_{a}}}{r^{2}\bar{\mu_{b}}}\right)
\end{equation}
\begin{equation}\label{Wt}
	\tilde{W}=-\tilde{U_{1}}=\frac{1}{2}\log\left(\dfrac{\bar{\mu_{b}}}{\bar{\mu_{0}}\bar{\mu_{c}}}\right),
\end{equation}
and since $\bar{U_{b}}-\bar{U_{a}}-\bar{U_{c}}=\bar{U_{b}}-\bar{U_{c}} -\tilde{U}-\log r-\bar{U_{b}}+\bar{U_{c}}=  -\tilde{U}-\log r$ we have
\begin{equation}\label{tildenu2}
	\tilde{\nu}=-\tilde{U}-\log r+\gamma_{00}+\gamma_{aa}+\gamma_{bb}+\gamma_{cc}-\gamma_{0a}+\gamma_{0c}-\gamma_{0b}-2\gamma_{bc}+\gamma_{ac}-\gamma_{ab}-\dfrac{1}{4}\log 2
\end{equation} We also have $ \tilde{V}=\tilde{\nu}-\tilde{U}-\tilde{W} $
\begin{equation}
	\tilde{V}=-2\tilde{U}-\tilde{W}-\log r+\gamma_{00}+\gamma_{aa}+\gamma_{bb}+\gamma_{cc}-\gamma_{0a}+\gamma_{0c}-\gamma_{0b}-2\gamma_{bc}+\gamma_{ac}-\gamma_{ab}-\dfrac{1}{4}\log 2\label{Vtilde}
\end{equation} 

We now determine the deformation field explicitly. The solution of Laplace equation is well known in cylindrical coordinates $(r,z)$.
As mentioned before, the functions $ \widehat{U} $ and $ \widehat{W} $ satisfy the Laplace equation (\ref{Laplace}). In the cylindrical coordinates the solution of Laplace equation is 
\begin{equation}\label{Xhat}
	\widehat{X}(r,z)=\sum_{n\ge 0}\left[A_{n}R^{n}+\hat{A}_{n}R^{-(n+1)}\right]P_{n}(\cos\vartheta)
\end{equation}
where
\begin{equation}
	R=\dfrac{\sqrt{r^{2}+z^{2}}}{m}, \quad \cos\vartheta=\dfrac{z}{R}
\end{equation}
$ P_{n}(\cos\vartheta) $ represent the Legendre polynomials of the first kind, $m$ is a scaling free parameter, and $ \widehat{X} $ refers to either $ \widehat{U} $ or $ \widehat{W} $.
The expansion (\ref{Xhat}) involves two sets of coefficients, $A_{n} $ and $ \hat{A}_{n} $, which can be attributed to interior and exterior multipole moments respectively\cite{Jackson}, in the sense that they represent distortions to the gravitational field that arise from the `near' and 'far' region respectively.  In this discussion, we only consider $ A_{n} $ coefficients, which describe the local distortion of a black black hole with a bubble due to external fields and will be called (interior) multipole moments. These deformations vanish as $R \to 0$ but diverge at infinity. 

It is possible to obtain the function $V$ by using equations  (\ref{VzVrho}) if the distortion fields $ \widehat{U} $ and $ \widehat{W} $ are already known. We can rewrite eq. (\ref{VzVrho}) as $\widehat{V} = \widehat{V}_1 +\widehat{V}_2$ where
\begin{equation}
	\widehat{V}_{1}=\widehat{V}_{\widehat{U}\widehat{U}}+\widehat{V}_{\widehat{W}\widehat{W}}+\widehat{V}_{\widehat{U}\widehat{W}}
\end{equation}
\begin{equation}\label{V2terms}
	\widehat{V}_{2}=\widehat{V}_{\tilde{U}\widehat{W}}+\widehat{V}_{\tilde{W}\widehat{U}}+2\widehat{V}_{\tilde{U}\widehat{U}}+2\widehat{V}_{\tilde{W}\widehat{W}}
\end{equation}
where 
\begin{equation}\label{Vfgr}
	V_{(fg),r}=r\left(f_{,r}g_{,r}-f_{,z}g_{,z}\right)
\end{equation}
\begin{equation}
	V_{(fg),z}=r\left(f_{,r}g_{,z}+f_{,z}g_{,r}\right)
\end{equation} The three parts $ \widehat{V}_{\widehat{U}\widehat{U}}, \widehat{V}_{\widehat{W}\widehat{W}}, \widehat{V}_{\widehat{U}\widehat{W}} $ involve only the distortion fields. We can write the function $ \widehat{V}_{1} $ in the following form:
\begin{equation}\label{V1}
	\widehat{V}_{1}=\sum_{n,k\ge 1}\dfrac{nk}{n+k}\left(a_{n}a_{k}+a_{n}b_{k}+b_{n}b_{k}\right)R^{n+k}\left[P_{n}P_{k}-P_{n-1}P_{k-1}\right]
\end{equation}
where $ P_{n}=P_{n}(z/R) $. 

Due to the fact that $ \widehat{U} $ and $ \widehat{W} $ are
logarithmic functions, we can further decompose each terms of \ref{V2terms} in the following manner:
\begin{equation}
\left(\widehat{V}_{\tilde{U}\widehat{X}}\right)	_{,r}=r\left(\tilde{U}_{,r}\widehat{X}_{,r}-\tilde{U}_{,z}\widehat{X}_{,z}\right)
\end{equation}
and we know
\begin{equation*}
	\tilde{U}=\bar{U_{c}}+\bar{U_{a}}-\bar{U_{b}}-\log r
\end{equation*}
By substituting it in the above equation we have
\begin{equation}
\left(\widehat{V}_{\tilde{U}\widehat{X}}\right)	_{,r}=r\left(\left(\bar{U_{c}}+\bar{U_{a}}-\bar{U_{b}}-\log r\right)_{,r}\widehat{X}_{,r}-\left(\bar{U_{c}}+\bar{U_{a}}-\bar{U_{b}}-\log r\right)_{,z}\widehat{X}_{,z}\right)~,
\end{equation}
\begin{eqnarray}
	\left(\widehat{V}_{\tilde{U}\widehat{X}}\right)	_{,r}=	\widehat{V}_{\bar{U_{c}}\widehat{X},r}+	\widehat{V}_{\bar{U_{a}}\widehat{X},r}-	\widehat{V}_{\bar{U_{b}}\widehat{X},r}-\widehat{X}_{r}~. 
\end{eqnarray}
Thus, 
\begin{equation}
		\widehat{V}_{\tilde{U}\widehat{X}}=	\widehat{V}_{\bar{U_{c}}\widehat{X}}+	\widehat{V}_{\bar{U_{a}}\widehat{X}}-	\widehat{V}_{\bar{U_{b}}\widehat{X}}-\widehat{X}~.\label{SumVWX2}
\end{equation}
Also $ \widehat{W}=\bar{U_{b}}-\bar{U_{0}}-\bar{U_{c}} $. By substituting it in \ref{Vfgr}, we get
\begin{equation}
\left(\widehat{V}_{\tilde{W}\widehat{X}}\right)_{,r}=r\left(\tilde{W}_{,r}\widehat{X}_{,r}-\tilde{W}_{,z}\widehat{X}_{,z}\right)
\end{equation}
\begin{eqnarray*}
	\begin{aligned}
	\left(\widehat{V}_{\tilde{W}\widehat{X}}\right)_{,r}&=r\left(\left(\bar{U_{b}}-\bar{U_{0}}-\bar{U_{c}}\right)_{,r}\widehat{X}_{,r}-\left(\bar{U_{b}}-\bar{U_{0}}-\bar{U_{c}}\right)_{,z}\widehat{X}_{,z}\right)\\
		&=r\left(\bar{U_{b}}_{,r}\widehat{X}_{,r}-\bar{U_{b}}_{,z}\widehat{X}_{,z}\right)-r\left(\bar{U_{0}}_{,r}\widehat{X}_{,r}-\bar{U_{0}}_{,z}\widehat{X}_{,z}\right)-r\left(\bar{U_{c}}_{,r}\widehat{X}_{,r}-\bar{U_{c}}_{,z}\widehat{X}_{,z}\right)\\
		&=\widehat{V}_{\bar{U_{b}}\widehat{X},r}-	\widehat{V}_{\bar{U_{0}}\widehat{X},r}-	\widehat{V}_{\bar{U_{c}}\widehat{X},r}
	\end{aligned}		
\end{eqnarray*}
\begin{equation}
\widehat{V}_{\tilde{W}\widehat{X}}=\widehat{V}_{\bar{U_{b}}\widehat{X}}-	\widehat{V}_{\bar{U_{0}}\widehat{X}}-	\widehat{V}_{\bar{U_{c}}\widehat{X}}\label{sumVWX}
\end{equation}
Then each term can be found by a line integral
\begin{equation}
	\widehat{V}_{\square\square}(r,z)=\int_{(r_{0},z_{0})}^{r,z} \left[ \left(	\widehat{V}_{\square\square}(r,z)\right)_{,r}\td r+\left(	\widehat{V}_{\square\square}(r,z)\right)_{,z}\td z\right]\label{LineIntegral}
\end{equation}
where the integral is taken along any path connecting the points $ (r_{0},z_{0}) $ and $ (r,z) $. Thus, the  field $ \widehat{V} $ is defined up to an arbitrary constant of integration defined by the point $(r_{0},z_{0}) $. We choose this arbitrary constant using a boundary condition. Here, $ \square $'s are to be filled with the corresponding notation for each term in $\widehat{V} $.
After calculating $ \widehat{V}_{\tilde{U}\widehat{U}} $, $ \widehat{V}_{\tilde{U}\widehat{W}} $, $ \widehat{V}_{\tilde{W}\widehat{U}} $ and $ \widehat{V}_{\tilde{W}\widehat{W}} $ we can obtain $ \widehat{V}_{2} $ by (\ref{V2terms}) for each of the multiple moments. 

Given the metric form (\ref{GWS distorted}), the functions (\ref{Ut}-\ref{Vtilde}) for the background spacetime, and (\ref{Xhat}) when $\hat{A}_n=0$, (\ref{V1}) for $\widehat{V}_1$, (\ref{V2terms}), equations (\ref{SumVWX2}), (\ref{sumVWX}), (\ref{LineIntegral}) for $\widehat{V}_2$, and $\widehat{V}=\widehat{V}_1+\widehat{V}_2$, we have the general form of the metric. The only terms that is given in terms of the integral is $\widehat{V}_2$ which is very tedious to compute it for higher orders. 

In this paper, we give explicit expressions for $n=1,2,3$ moments (here $A_n, B_n$ denote the coefficients in the expansions of $\widehat{U}, \widehat{W}$ respectively): 
\begin{equation}\label{V2n1}
	\widehat{V}_{2}(r,z)=(\frac{-1}{2m})\left[(B_{1}-A_{1})(R_{b}-R_{c})+(2A_{1}+B_{1})R_{a}-(A_{1}+2B_{1})R_{0}+3(A_{1}+B_{1})z\right] \quad, n=1
\end{equation}
\begin{equation}\label{v2n2}
	\begin{aligned}
		\widehat{V}_{2}(r,z)&=\left(\frac{-1}{2m^{2}}\right)\left[(2A_{2}+B_{2})(a+z)R_{a}+(b+z)(B_{2}-A_{2})R_{b}-(A_{2}+2B_{2})zR_{0}+(c+z)(A_{2}-B_{2})R_{c}\right]\\
		&-\left(\frac{-1}{2m^{2}}\right)\left(\frac{3}{2}(A_{2}+B_{2})(r^{2}-2z^{2})\right)\quad, n=2
	\end{aligned}
\end{equation}

\begin{equation}\label{v2n3}
	\begin{aligned}
		\widehat{V}_{2}(r,z)&=	\left(\frac{-1}{4m^{3}}\right)\left[(r^{2}-2z^{2})(2B_{3}+A_{3})R_{0}+(2a^{2}+2z^{2}-r^{2}+2az)(B_{3}+2A_{3})R_{a}\right]
		\\
		&-\left(\frac{1}{4m^{3}}\right)\left[(2b^{2}+2z^{2}-r^{2}+2bz)(B_{3}-A_{3})R_{b}-(2c^{2}+2z^{2}-r^{2}+2cz)(B_{3}-A_{3})R_{c}\right]\\
		&-\left(\frac{1}{4m^{3}}\right)\left[(6z^{3}-9r^{2}z)(A_{3}+B_{3})\right]\quad, n=3
	\end{aligned}
\end{equation}
Note that for all of these $ \widehat{V}=\widehat{V}_1+\widehat{V}_2$ satisfy the eqs. \ref{VzVrho}, which implies vacuum Einstein equations are satisfied. However, we have also explicitly confirmed that metric (\ref{GWS distorted}) with the functions (\ref{Ut}-\ref{Vtilde}) for the background spacetime, (\ref{Xhat}) for $\widehat{U}$ or $\widehat{W}$ and (\ref{V1}) for $\widehat{V}_1$ and $\widehat{V}_2$ given by (\ref{V2n1}-\ref{v2n3}) satisfy Einstein equations. Note that for $\widehat{V}_2$, we can choose any value of $n$ or sum of any combinations. The same is true for $\widehat{U}$, $\widehat{W}$, or $\widehat{V}_1$. However, in analogy to the multiple expansion in electromagnetism or gravitational Newtonian potentials of external sources (corresponding to the interior multiple moments), $A_n$ and $B_n$ become weaker with increasing value of $n$. In 4-dimensional case these relativistic multiple moments were mapped to Newtonian multiple moments of a ring of mass or two masses along the axis in \cite{Abdolrahimi2015B} in the simplest case.  

\subsection{Regularity} As a result of the constraint \eqref{deformconstraint}, the rod structure, which is determined by the zeroes of the determinant of the restriction of the metric to the Killing fields (in the Weyl coordinate system, this is simply the function $r^2$), is not changed by the deformation. The deformations are characterized by harmonic functions which diverge at asymptotic infinity but remian smooth and bounded everywhere else. Analyticity of the metric is thus guaranteed, as in the undistorted case, away from the singular sets of the harmonic functions. We have also confirmed the regualrity of the spacetime by explicit computation of the curvature invaraints of the spacetime. 
\subsubsection{Dipole deformations}
The dipole term in a multipole expansion represents the lowest-order contribution to the field or potential, followed by the quadrupole, octupole, and higher-order terms. The strength of these terms diminishes with increasing order, meaning that the dipole term typically dominates the behaviour of the system compared to higher-order terms such as the quadruple. In many physical systems, especially those with a significant degree of symmetry, the dipole term may be the most important in determining the overall behaviour. In what follows a dipole deformation refers to the case where $A_1\neq0$, $B_1\neq0$, and $A_{n\geq 1}=B_{n\geq 1}=0$. Similarly, the quadruple case refers to when $A_2\neq0$, $B_2\neq0$, and $A_{n\neq 2}=B_{n\neq 2}=0$. The functions $\hat{U}$ and $\hat{W}$ and thus multiple moments $A_n$ and $B_n$ are independent, and correspond to independent distortion fields. 

When addressing the conical singularities using multiple moments, the goal is to utilize terms such as dipole and/or quadruple to remove the conical singularities from the local black hole bubble solution. Although in removing the conical singularities we would need to impose conditions on values of multiple moments. 

Let us consider the solution on the negative semi-infinite rod $z < 0$, 
\begin{equation}
\beta_1 = \lim_{r \to 0} \frac{r^2 e^{\nu}}{e^{2U_1}} = \exp\left[-(2A_1 +B_1)a - (A_1 - B_1)(c-b) \right]
\end{equation} whereas for the finite bubble rod with $b < z < c$, 
\begin{equation}
\beta_2 = \lim_{r \to 0} \frac{r^2 e^{\nu}}{e^{2U_1}} =  \frac{(c-b)^2}{c(c-a)} \exp \left[ (2A_1 + B_1)a - (A_1 - B_1)(c+b) \right]
\end{equation} Regularity requires $\beta_1 = \beta_2$ and then the period of $\psi$ must be $\Delta \psi = 2\pi \sqrt{\beta_1}$.  On the semi infinite rod $ z > c$, the Killing vector field $\partial_\phi$ degenerates, with regularity requiring
\begin{equation}
\beta_3 = \exp\left[ (2A_1 + B_1)a + (A_1 - B_1)(c-b) \right]
\end{equation} Similarly on the finite rod $a < z < b$, 
\begin{equation}
\beta_4 = \frac{(b-a)b}{c(c-a)} \exp\left[ (2 A_1 + B_1)a - (A_1 - B_1) (c-b) \right]
\end{equation} Regularity requires $\beta_3 = \beta_4$ and that the period of $\phi$ must be $\Delta \phi = 2\pi \sqrt{\beta_3}$.  Then
\begin{equation}
\frac{\beta_3}{\beta_4} = 1 = \frac{(c-a) c}{(b-a) b} \exp\left[ 2(A_1 - B_1)(c-b)\right]
\end{equation} and 
\begin{equation}
\frac{\beta_2}{\beta_1} = 1 = \frac{(c-b)^2}{c(c-a)} \exp\left[2a (2A_1 + B_1) - 2b(A_1 - B_1)\right]
\end{equation} The former gives
\begin{equation}\label{regB0}
e^{B_1} = \left(\frac{c(c-a)}{b(b-a)}\right)^{\frac{1}{2(c-b)}} e^{A_1}
\end{equation} which leads to
\begin{equation}\label{regA1}
e^{A_1} = \left[ \frac{c(c-a)}{(c-b)^2} \left(\frac{b(b-a)}{c(c-a)}\right)^{\frac{b+a}{c-b}}\right]^{\frac{1}{6a}}
\end{equation}  Substituting this back into \eqref{regB0} gives
\begin{equation}\label{regB1}
e^{B_1} =\frac{ (c(c-a))^{\frac{2a -2b + c}{6a(c-b)}} (b(b-a))^{\frac{b-2a}{6a(c-b)}}}{(c-b)^{^{\frac{1}{3a}}}}.
\end{equation} Note that the right hand sides of \eqref{regA1} and \eqref{regB1} are strictly positive so that $A_1$ and $B_1$ are uniquely defined. Of course, as the rod points of the undistorted solution approach each other (i.e. $a \to b$ or $b \to c$), the distortions required to achieve regularity become arbitrarily large. We consider such case unrealistic.

In summary, regularity of the metric is achieved provided we choose the distortion parameters $(A_1, B_1)$ to satisfy \eqref{regA1} and \eqref{regB1} respectively. In this case the periods of $\psi$ and $\phi$ must be chosen to be
\begin{equation}\label{Dpsiphin1}
    \Delta\psi = 2\pi \sqrt{\beta_1},\quad \Delta \phi = 2\pi \sqrt{\beta_3}
\end{equation}

 where
\begin{equation}\label{beta1beta3n1}
\beta_1 = (e^{A_1})^{-(2a + c - b)} (e^{B_1})^{c-b -a}, \qquad \beta_3 = (e^{A_1})^{2a + c-b} (e^{B_1})^{a - (c-b)}.
\end{equation} 
Near the horizon surface, the dipole distorted metric is
\begin{eqnarray*}
    ds^2&&=-\exp\left[2(A_{1} + B_{1})z\right]\left(\frac{1}{4z(a-z)}r^2-\frac{(a^2-2az+2z^2)}{16z^3(a-z)^3}r^4+O(r^6)\right) \td t^2 \\
&&+\exp{\left[(-2a + b - c + 2z)A_{1} -(a + b - c - 2z)B_{1}\right]}\left(\frac{ab }{2c z(a - z)}+O(r^2)\right) (\td r^2 + \td z^2)\\
&&+\left(\left(\frac{2z(-z + b)}{(c - z)}+\frac{(z^4 - 4cz^3 + (4bc + 2c^2)z^2 - 2bc(b + c)z + b^2c^2)}{2z(c - z)^3(-z + b)}r^2\right) \exp(-2B_{1}z)+O(r^4) \right) \td \psi^2 \\
&& \left( \frac{2(a - z)(c - z)\exp(-2A_{1}z)}{(-z + b)}\right) \td \phi^2
\end{eqnarray*}
The metric on the horizon of the distorted black hole is given by
\begin{align}
    \td s^2_3
   &=2a\left(\exp\left[((A_1 + B_1)(2\cos^2\theta-1) - A_1)a + (A_1 - B_1)(b - c)\right]\frac{b}{c}\td\theta^2 \right.\\
   &+ \left.\exp[-2a B_1\cos^2\theta]\frac{(b-a\cos^2\theta)\cos^2\theta}{(c-a\cos^2\theta)}\td\psi^2 +\exp[-2aA_1\cos^2\theta]\frac{(c-a\cos^2\theta)\sin^2\theta}{(b-a\cos^2\theta)}\td\phi^2\right) \nonumber
    \end{align} where $0 < \theta < \pi/2$ and $z = a/2 + a/2 \cos\theta$. The metric extends to a smooth metric on $S^3$. Note that removing the conical singularity at $\theta =\pi/2$ requires that the period of $\psi$ is given by $\Delta \psi = 2\pi\sqrt{\beta_3}$. This corresponds to the the removal of the conical singularity on the semi infinite rod $z < 0$. Similarly, removing the conical singularity at $\theta=0$ requires that the period of $\phi$ is given by $\Delta \phi = 2\pi \sqrt{\beta_4}$. This corresponds to removing the conical singularity on the disc $a < z < b$.

\subsubsection{Quadropole deformations}
We now turn to the $n=2$ deformations parameterized by $(A_2, B_2)$. For the quadruple case, we consider $A_1=B_1=0$. We seek to mitigate conical singularities by choosing appropriate values of multiple moments of $(A_2, B_2)$. Such adjustments show that we can balance the black hole bubble system when the black hole is surrounded by appropriate sources.

On the negative semi-infinite rod $z < 0$, 
\begin{equation}
\gamma_1 = \lim_{r \to 0} \frac{r^2 e^{\nu}}{e^{2U_1}} = \exp\left[-(2A_2 + B_2)a^2 - (A_2 - B_2) (c^2 - b^2) \right]
\end{equation}  whereas for the finite rod with $b < z < c$, 
\begin{equation}
\gamma_2 = \lim_{r \to 0} \frac{r^2 e^{\nu}}{e^{2U_1}} = \frac{(c-b)^2}{c(c-a)} \exp \left[(2A_2 + B_2)a^2 - (A_2 - B_2) (c^2 + b^2) \right]
\end{equation} Regularity requires $\gamma_1 = \gamma_2$ and that we impose the periodicity condition $\psi \sim \psi + 2\pi \sqrt{\gamma_1}$. On the semi infinite rod $ z > c$, the Killing vector field $\partial_\phi$ degenerates, with regularity requiring
\begin{equation}
\gamma_3 = \exp\left[ (2A_2 + B_2)a^2 + (A_2 - B_2)(c^2 - b^2)  \right]
\end{equation} Similarly on the finite rod $a < z < b$, 
\begin{equation}
\gamma_4 = \frac{(b-a)b}{c(c-a)} \exp\left[ (2 A_2 + B_2)a^2 - (A_2 - B_2) (c^2-b^2) \right]
\end{equation} Regularity requires $\gamma_3 = \gamma_4$ and that the period of $\phi$ must be $\Delta \phi = 2\pi \sqrt{\gamma_3}$.  Using $\gamma_3 / \gamma_4 = 1$ implies
\begin{equation}
e^{B_2} = \left[\frac{c(c-a)}{b(b-a)}\right]^{\frac{1}{2(c^2-b^2)}} e^{A_2}
\end{equation} and using $\gamma_2/\gamma_1 = 1$ fixes
\begin{equation}
e^{A_2} = \left[ \frac{(b(b-a))^{\frac{a^2 + b^2}{c^2-b^2}}}{(c-b)^2} \left(c(c-a)\right)^{\frac{-a^2 - 2b^2  + c^2}{c^2-b^2}}\right]^{\frac{1}{6a^2}}
\end{equation} which gives
\begin{equation}
e^{B_2} = \left(\frac{1}{(c-b)}\right)^{\frac{1}{3a^2}}\left[b(b-a)\right]^{\frac{b^2 - 2a^2}{6a^2(c^2-b^2)}} \left[c(c-a)\right]^{\frac{2a^2-2b^2 + c^2}{6a^2(c^2-b^2)}}
\end{equation} In summary, we must select the period of $\psi$ to be
\begin{equation}
\Delta \psi = 2\pi \sqrt{\gamma_1}, \qquad \gamma_1 = e^{-(2a^2 + c^2 - b^2)A_2} e^{(c^2 -a^2 - b^2) B_2}
\end{equation} and the period of $\phi$ must be chosen such that
\begin{equation}
\Delta \phi = 2\pi \sqrt{\gamma_3}, \qquad \gamma_3 = e^{(2a^2 + c^2-b^2)A_2} e^{(a^2 - c^2 + b^2)B_2}
\end{equation}
 The horizon metric of the deformed black hole in terms of the quadropole parameters $A_2$ and $B_2$ is given by
 \begin{align}
     \td s^2_3 &=\frac{2ab}{c} \exp[-a^2(2A_2+B_2) - (A_2 - B_2)(c^2-b^2) + 2a^2(A_2 + B_2) \cos^4\theta] \nonumber\\ &+\frac{2a \exp[-2a^2 B_2 \cos^4\theta] (b - a\cos^2\theta) \cos^2\theta}{c - a \cos^2\theta} \td\psi^2 \nonumber\\ &+ \frac{2a \exp[-2a^2 A_2 \cos^4\theta] (c - a\cos^2\theta) \sin^2\theta}{b - a \cos^2\theta} \td\phi^2
 \end{align}
where $0<\theta<\pi/2$ and $z=a/2(1+\cos 2\theta)$. 
 Where, this also confirms that to have a smooth metric on $S^3$, we have the above regularity conditions $\Delta \psi = 2\pi \sqrt{\gamma_1}$ and $\Delta \phi = 2\pi \sqrt{\gamma_4}$.
\begin{figure}
	\centering
	\includegraphics[width=10cm]{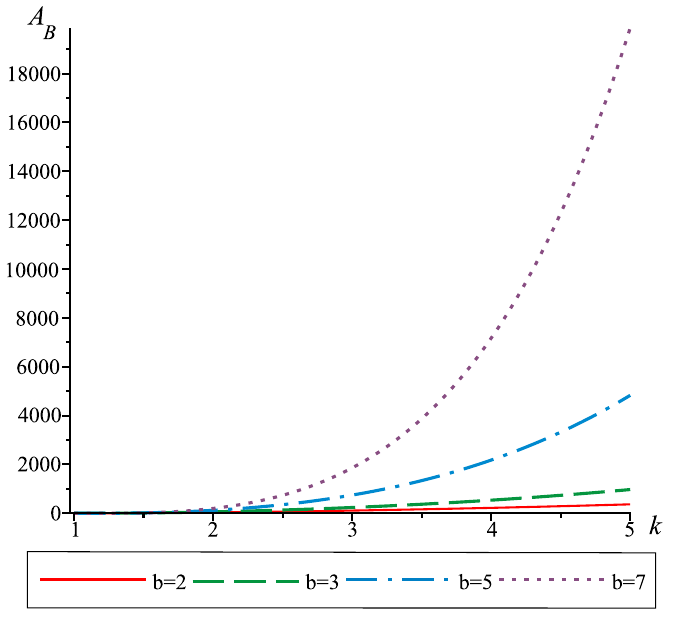}
		\caption{Area of the bubble $A_{B}$ for dipole distortions of the DBHB for several values of b. In this plot $a=1$, $c=kb$.% and only the dipole multipole moment case is considered.
  }
	\label{ABn1}
\end{figure}
%\begin{figure}
%	\centering
	%\includegraphics[width=10cm]{ UndisABn1_flat.pdf}
		%\caption{The behaviour of the area of the bubble, $A_{B}$, for %the AF black hole-bubble solution for values of b. In this plot %$a=1$ and $c=kb$.}
	%\label{UNABn1F}
 %\end{figure}
\begin{figure}
   \centering
   \includegraphics[width=8cm]{ 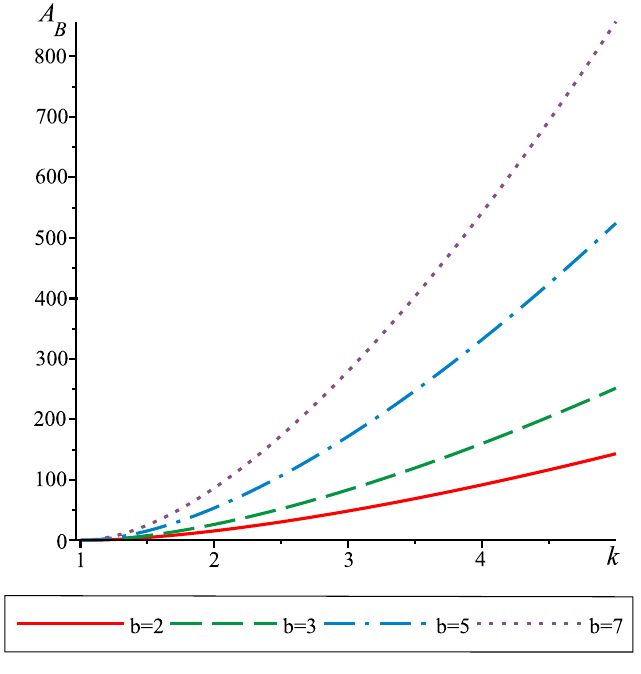}
	\includegraphics[width=8cm]{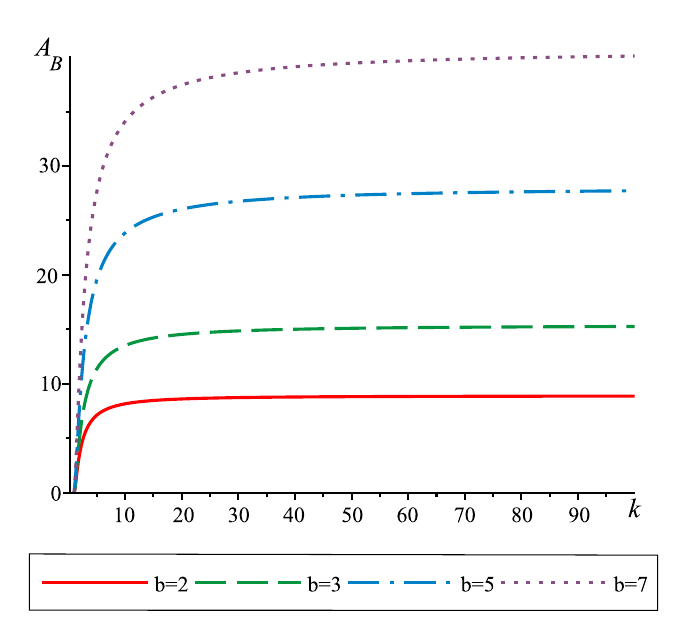}
		\caption{Area of the bubble $A_B$ for dipole distortions of the BHB-AF solution (left) and the BHB-NF solution (right) for several values of $b$. 
  %The behaviour of the area of the bubble, $A_{B}$, for the BHB-AF solution and BHB-NF solution on the right and left, respectively. 
  For both cases %In this plot,
  $a=1$, $c=kb$.}
	\label{UNABn1NF}
\end{figure}
\begin{figure}
    \centering
	\includegraphics[width=11cm]{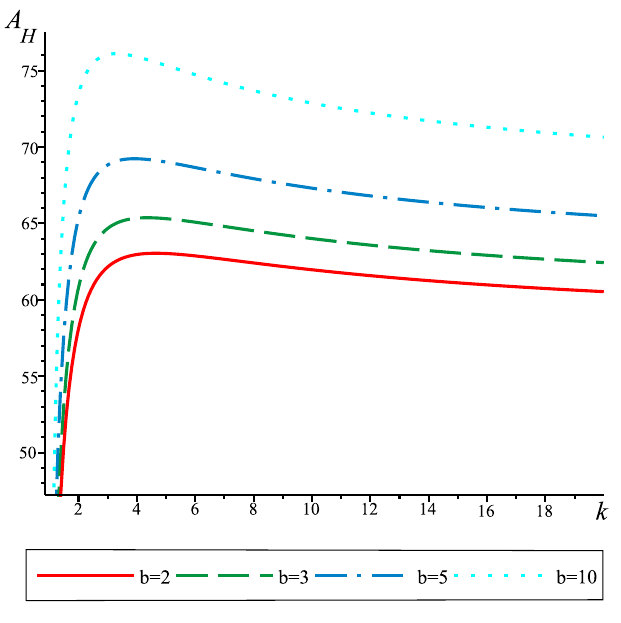}
	\caption{Area of the horizon $A_H$ for dipole distortions of the DBHB for different values of $b$. 
 %The behaviour of area of the horizon $A_{H}$ for the DBHB for different values of b. 
 In this plot, $a=1$ and $c=kb$. %and the dipole multiple moment case is considered.
 }
	\label{DAHB1}
 \end{figure}
\begin{figure}
 \centering
 \includegraphics[width=8cm]{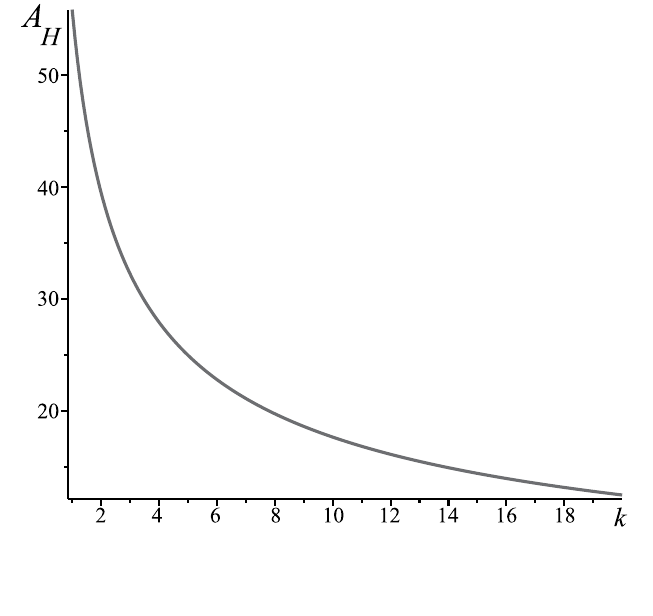}
\includegraphics[width=8cm]{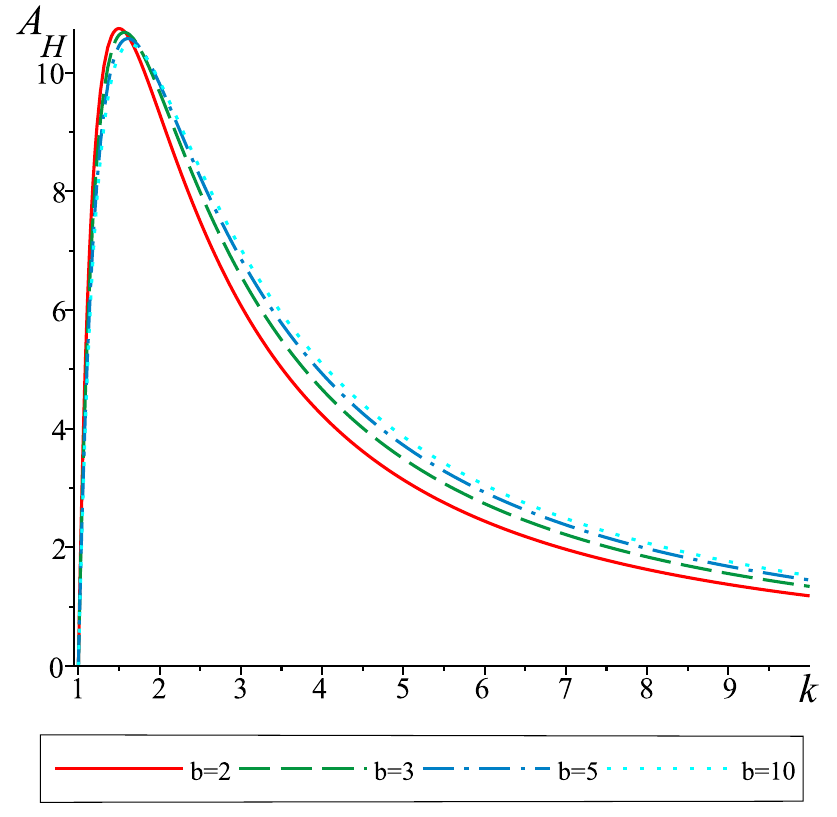}
	\caption{Left side: area of the horizon, $A_{H}$, for dipole distortions of AF black hole-bubble solution for different values of b. In this plot $a=1$ and $c=kb$. Right side: Area of the horizon, $A_{H}$, for dipole distortions of the BHB-nf $b$. In this plot $a=1$ and $c=kb$. }
	\label{UnAHB1NF}
\end{figure}
\begin{figure}
    \centering
	\includegraphics[width=11cm]{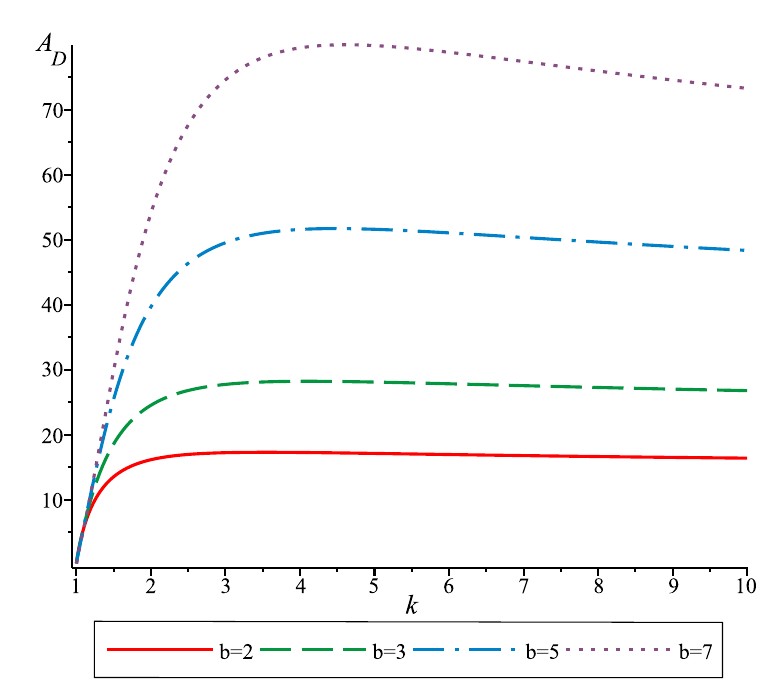}
	\caption{Area of the disk, $A_{D}$, for dipole distortions of DBHB for several values of b. In this plot, $a=1$, $c=kb$.}
	\label{DADB1}
 \end{figure}
 %\begin{figure}
% \centering
%	\includegraphics[width=11cm]{UndisADflat.pdf}
%	\caption{The behavior of the area of the disk, $A_{D}$, for the AF black hole bubble solution. In this plot $a=1$ and $c=kb$.}
	%\label{UnADB1F}
% \end{figure}
\begin{figure}
 \centering
 \includegraphics[width=9cm]{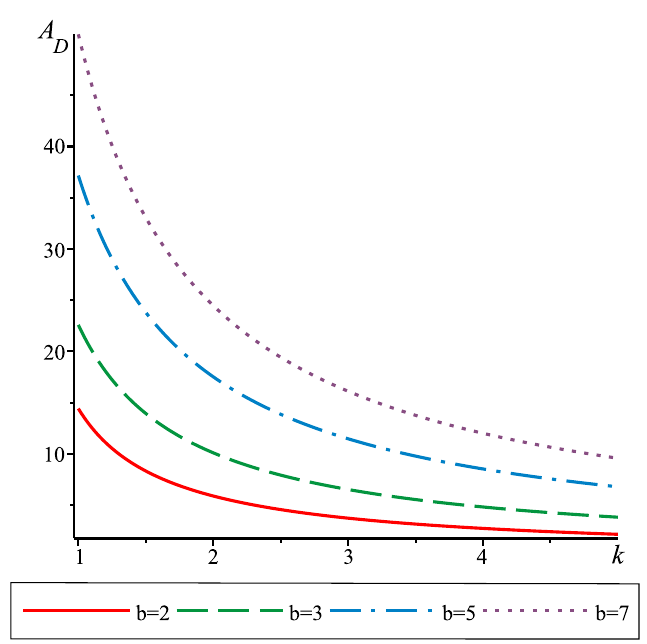}~~~
	\includegraphics[width=9cm]{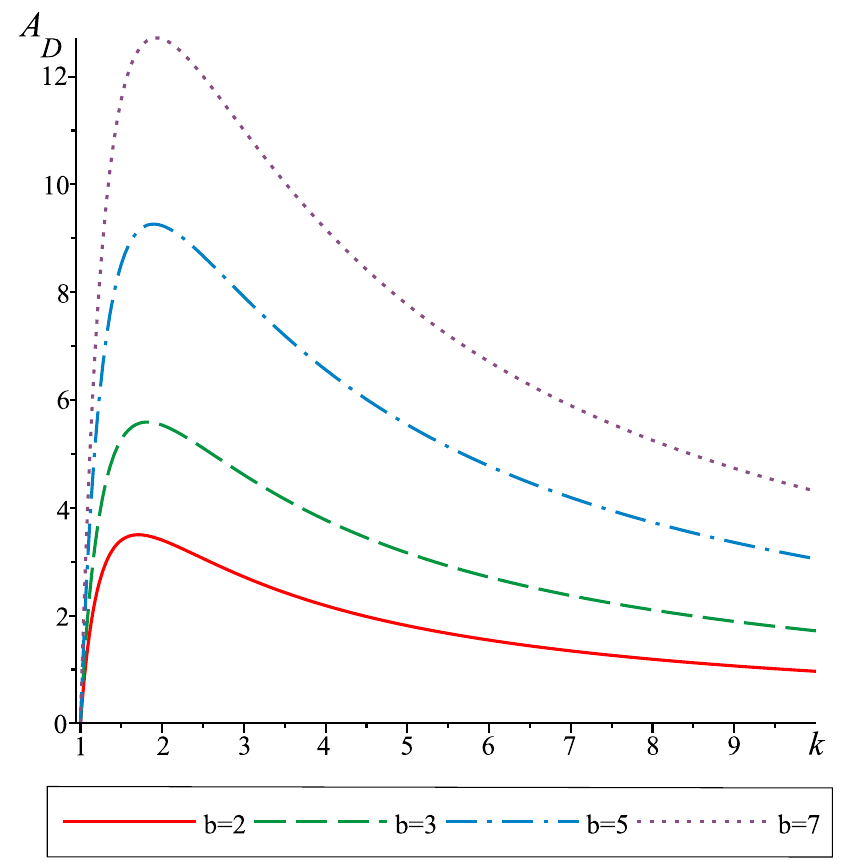}
	\caption{Left side: area of the disk, $A_{D}$, for dipole distortions of the AF black hole bubble solution. In this plot $a=1$ and $c=kb$. Right side: area of the disk, $A_{D}$, for dipole distorted BHB-nf for different values of $b$. In this plot $a=1$ and $c=kb$.}
	\label{UnADB1NF}
\end{figure}

%\begin{figure}
% \centering
	%\includegraphics[width=11cm]{EtaBn1cbCONSTANT.pdf}
	%\caption{The behaviour of $\eta_{B}$=(area of the distorted %bubble)/(area of the AF undistorted bubble) for different values of %$\delta$.  In each plot, $c-b=\delta$ is a constant and a=1.}
	%\label{EtaBb}
%\end{figure}
\begin{figure}
 \centering
 \includegraphics[width=8cm]{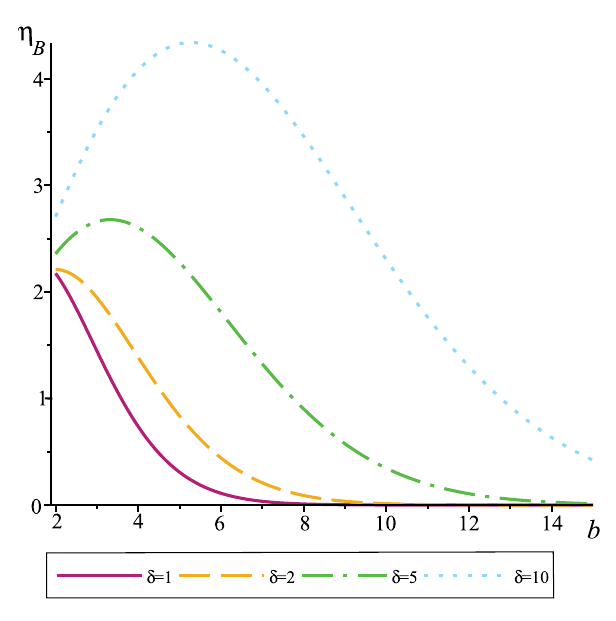}~~~
	\includegraphics[width=8cm]{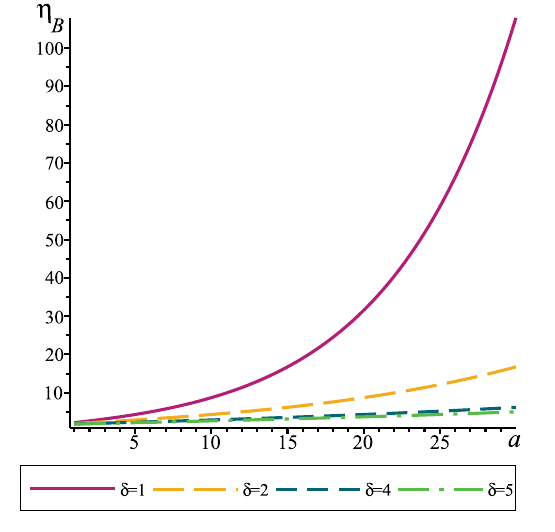}
	\caption{Left side: $\eta_{B}$=(area of the distorted bubble)/(area of the AF undistorted bubble) for different values of $\delta$.  In the first plot, $c-b=\delta$ is a constant and $a=1$. Right side: $\eta_{B}$=(area of the distorted bubble)/(area of the undistorted AF bubble) for different values of $\delta$.  In each plot, both $c-b=\delta$ and  $b-a=\delta$ are constant.}
	\label{EtaBb2}
\end{figure}

%\begin{figure}
 %\centering
	%\includegraphics[width=11cm]{EtaHn1cbCONSTANT.pdf}
	%\caption{The behavior of $\eta_{H}$=(The area of the distorted %horizon)/(The area of the horizon of the AF undistorted solution) %for different values of $\delta$.  In each plot, $c-b=\delta$ is a %constant. Here $a=1$}
	%\label{EtaHb}
%\end{figure}
\begin{figure}
 \centering
 \includegraphics[width=8cm]{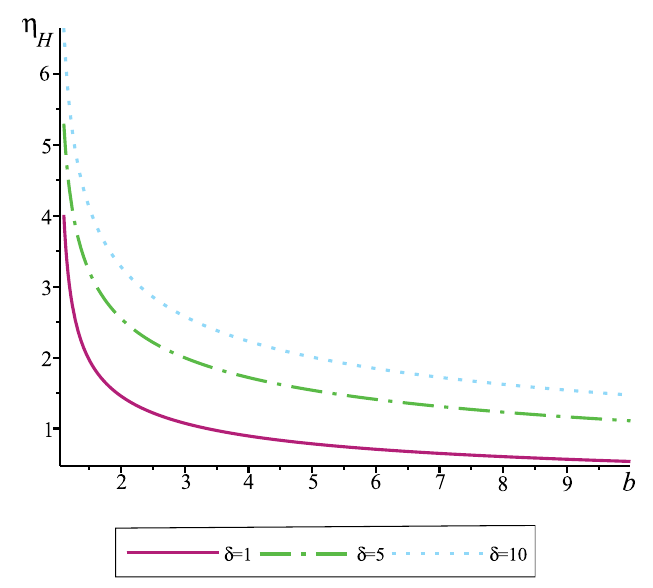}~~~
	\includegraphics[width=8cm]{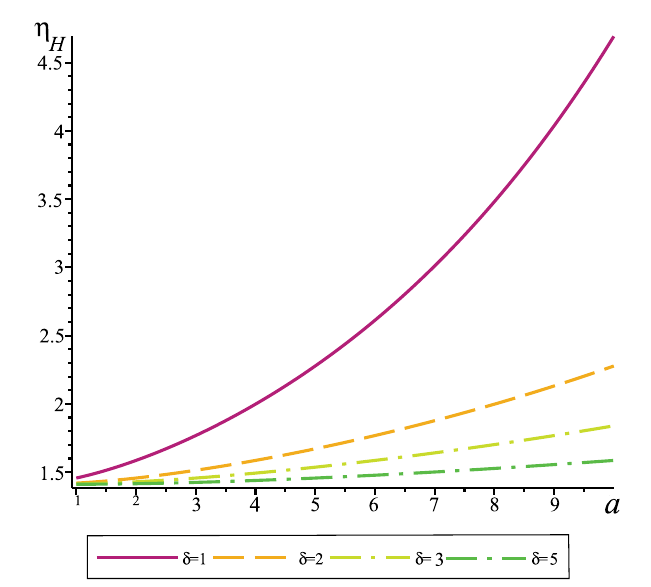}
	\caption{Left side: $\eta_{H}$=(The area of the distorted horizon)/(The area of the horizon of the AF undistorted solution) for different values of $\delta$.  In each plot, $c-b=\delta$ is a constant. Here $a=1$. Right side: $\eta_{H}$=(area of the distorted horizon)/(area of the AF undistorted horizon) for different values of $\delta$.  In each plot, both $c-b=\delta$ and  $b-a=\delta$  are constant.}
	\label{EtaHb}
\end{figure}
\subsection{Physical properties} 
In this section, we discuss various classical properties of the black hole, bubble, and disk. Simple expressions are obtained for horizon/bubble
area, and surface gravity.
We find that the Komar integral over the horizon \eqref{Komar} is unchanged when the black hole gets distorted. Recall that the surface gravity is defined by 
\begin{equation}
\kappa^2=-\frac{1}{2}(\nabla^\alpha \xi^\beta)(\nabla_\alpha \xi_\beta),
\end{equation}
where, $\nabla_\alpha$ is the covariant derivative defined with respect to the metric and limit as the horizon is approached is understood.  Here, we have considered $\xi^\alpha_{(t)}=\delta^\alpha_t$. However, since the spacetime is not asymptotically flat, the surface gravity is defined only up to an arbitrary ``red-shift" factor. More precisely, the surface gravity is only defined up to an arbitrary constant which depends on the normalization of the time-like Killing vector. In this case, the quantity of the surface gravity is changed by an exponential factor (which can be absorbed considering the ``red-shift" factor),  
\begin{equation}
\kappa = \lim_{r\to 0} \sqrt{\frac{e^{2U_0}}{r^2 e^{2\nu}}} = \begin{cases}  \sqrt{\frac{c}{2ab}} \exp\left[\frac{1}{2} \left((2a + c-b)A_1 + B_1(a - c + b)\right)\right] &\text{ if }\;  n=1 \\
\sqrt{\frac{c}{2ab}}  \exp\left[\frac{1}{2} \left((2a^2 + c^2-b^2)A_2 + (a^2 - c^2 + b^2)B_1\right)\right] &\text{ if }\;  n=2 . 
\end{cases}
\end{equation} 
The surface area of a spatial cross-section of the black hole horizon is a key feature that gives its entropy and information content. For the undistorted black holes, it obeys a `second law', or Area Theorem, subject to energy conditions, that requires it to be non-decreasing. 
Note that, in general a distorted and an undisturbed black hole are not related to each other by a dynamical process of bringing masses around a black hole adiabatically from infinity, although such process could be formally imagined. 
The area of the horizon for the dipole distortion is
\begin{equation}
A_H =  \Delta\phi \Delta \psi  \sqrt{\frac{2 a^3 b}{c}} \exp\left[\frac{1}{2}\left(-(2a + c - b)A_1 + (c - a - b)B_1\right)\right]~,
\end{equation} and for the quadruple distortion it is
\begin{equation}
A_H =  \Delta\phi \Delta \psi  \sqrt{\frac{2 a^3 b}{c}} \exp\left[\frac{1}{2}\left(-(2a^2 + c^2 - b^2)A_2 + (c^2 - a^2 - b^2)B_2\right)\right]~.
\end{equation} We could  formally define the mass of the non-asymptotically flat solution by setting 
\begin{equation}
m=\frac{3}{16\pi}\kappa A_H~. \label{SmarUDBH}
\end{equation} although, we emphasize that this identification does not come from some well-defined ADM energy. 
As is well known, mechanical laws of black holes represent relations between the black hole variables, such as mass, horizon area, surface gravity. These can be thought of as describing the tangent space within a family of solutions (i.e. a linearization). The classic work of Bekenstein \cite{Bekenstein:1973ur}, Bardeen, Carter, and Hawking \cite{BCH} has shown that the mechanical laws
governing classical systems containing black holes can be
placed in analogy with those of thermodynamics. This correspondence has been made explicit for many solutions. 

Geroch and Hartle investigated this correspondence for the systems of black hole plus distorting matter for a Schwarzschild black hole in four dimensions \cite{GerochH}. They considered the black hole as a single system acted upon by the gravitational
forces of the external matter and found that its laws continue
to have a simple correspondence with those of
thermodynamics. The first law of black hole mechanics delineates a connection between two equilibrium states of a black hole, altered by variations in its mass, horizon area, and additional parameters such as electric charge and angular momentum, alongside alterations in the stress-energy of surrounding matter, if applicable. The global first law applies to the system comprising both the black hole and the external matter influencing it. Extending spacetime to attain asymptotic flatness is necessary to define the global first law of black hole mechanics. Achieving this extension is done under the assumption that the solution can incorporate the sources of the distorting matter. To achieve such an extension, we need to violate the Einstein equations within the vicinity of the source's position. Then, such an extension requires considering the solution at the exterior region of sources and requiring the decay of the distortion field as we approach spatial infinity and extending the manifold to infinity. 

The constructed solution in this paper is valid only in the interior region of the sources. Only when such an extension is formally performed beyond the region where external sources are located, can we then normalize the timelike Killing vector $\bf{\xi}_{(t)}$ at spatial infinity such that $\bf{\xi}^2_{(t)}$. 
On the other hand, the local first law applies specifically to the black hole system itself, excluding any consideration of distorting matter within black hole mechanics. It is formulated by observers residing close to the black hole who solely attribute the local gravitational field to the black hole, viewing it as an isolated and undistorted entity. Thus, assuming that there is no other matter presenting the space-time is asymptotically flat, they define its surface gravity, the horizon area, and the black hole Komar mass such that they satisfy the Smarr formula \ref{SmarUDBH}. Note that as far as the Smarr relation is concerned, the bubble and disc surface will not contribute in the vacuum, although such contributions do appear when refinements to the Smarr relation are considered, such as those discussed for asymptotically flat vacuum solutions in \cite{Kunduri:2018qqt}.

The area of the bubble and disc of the distorted solutions are, for $n=1$
\begin{equation}\begin{aligned}
A_B &=  \Delta \phi \frac{c-b}{\sqrt{c(c-a)}}\int_b^c \sqrt{\frac{z}{z-a}}\exp\left[\frac{1}{2} \left((2a - b - c - 2z)A_1 + (a + b + c - 2z)B_1\right)\right] \; \td z \\
A_D & =  \Delta\psi \sqrt{\frac{b(b-a)}{c(c-a)}} \int_a^b\sqrt{\frac{z}{z-a}} \exp\left[\frac{1}{2} \left((2a + b - c - 2z)A_1 + (a - b + c - 2z)B_1\right)\right] \; \td z.
\end{aligned} 
\end{equation} and for the quadruple distortion, 
\begin{equation}
\begin{aligned}
A_B & =  \frac{\Delta \phi(c-b)}{\sqrt{c(c-a)}}\int_b^c \sqrt{\frac{z}{z-a}}\exp\left[\frac{1}{2} \left((2a^2 - b^2 - c^2 - 2z^2)A_2 + (a^2 + b^2 + c^2 - 2z^2)B_2\right)\right] \; \td z \\
A_D & =  \Delta \psi \sqrt{\frac{b(b-a)}{c(c-a)}} \int_a^b \sqrt{\frac{z}{z-a}} \exp\left[\frac{1}{2} \left((2a^2 + b^2 - c^2 - 2z^2)A_2 + (a^2 - b^2 + c^2 - 2z^2)B_2 \right)\right] \; \td z.
\end{aligned}
\end{equation}  
In this section, we consider the area of the black hole, bubble, and disk for the smooth distorted black hole bubble solution (DBHB), i.e., where the distortion parameters $A_i, B_i$ are chosen to remove the conical singularity. This requires that the period of $\psi$ and $\phi$ are given by (\ref{Dpsiphin1}). We consider the area of the black hole horizon, bubble and disk for the asymptotically flat undistorbed black hole bubble solution, where $\psi$ and $\phi$ are identified with period $2\pi$. We will, for brevity, sometimes refer to this solution as BHB-AF, since the spacetime is asymptotically flat, although it is not free of conical singularities in the interior region. As a comparison, also consider the area of the black hole, bubble and disk for the undistorted black hole-bubble solution, where the period of $\phi$ and $\psi$ are given by $\Delta\phi=2\pi\sqrt{b(b-a)/c(c-a)}$ and $\Delta\psi=2\pi(c-b)\sqrt{1/c(c-a)}$, respectively. We will denote this solution by BHB-NF, since the spacetime is not asymptotically flat. In this case, the bubble and disc are smooth, although the horizon $S^3$ will have conical singularities at one of its poles ($z=0$).

In these various cases, the solutions are parameterized by the positive numbers $0<a<b<c$. In order to simplify the analysis, we will use the scaling freedom to arrange $a=1$. We can then consider plots of physical quantities with $c = k b$ where $k>1$ while varying the distance $b-a$. Ideally, one would like to eliminate these parameters in favour of physical quantities, namely the area of the horizon, bubble, and disc. One could then compare, e.g. the size of the horizons of the distorted and undistorted cases with fixed bubble area. However, we have found it too complicated to perform this explicitly. It should be noted, of course, that the interval lengths $b-a$ and $c-b$ are, in fact, geometric invariants that can be used to uniquely characterize vacuum Weyl solutions~\cite{Hollands:2007aj}.

In Figure \ref{ABn1}, we consider the area of the bubble $A_{B}$ for DBHB, for different values of $b$ in the case of dipole deformations when we increase $c=k b$. We compare this area to the case of the BHB-AF or to BHB-NF in Figure \ref{UNABn1NF}. Roughly, as $b$ is increased, the horizon area decreases while the bubble and disc areas grow.  Qualitatively, the BHB-AF and DBHB show similar behaviour. As $k$ is varied, we expect the area of the bubble to grow (indeed as $k\to 1$ the bubble rod disappears). Consider next figure \ref{DAHB1}, in which plots the area of the horizon $A_{H}$ for DBHB, for different values of $b$ in the case of smooth dipole deformations. We compare this area to the case of the BHB-AF and BHB-NF in Fig. \ref{UnAHB1NF}. Here, the distorted black hole-bubble shows qualitatively different behaviour; as $k$ grows (the size of the disc rod becomes large) the area of the horizon approaches a constant independent of $k$, whereas in the undistorted case, the area of the horizon grows arbitrarily small. In particular, for the BHB-AF case, the horizon area is independent of $b$. 
In Fig \ref{DADB1}, we consider the area of the disk $A_{D}$ for the DBHB, for different values of $b$ in the case of dipole deformations. We compare this area to the case of BHB-AF and BHB-NF in Fig. \ref{UnADB1NF}. 
We observe that for BHB-AF when we increase $c$, the area of the bubble increases while the area of the black hole horizon and the area of the disk both decrease (see figures \ref{UNABn1NF},\ref{UnAHB1NF}, and \ref{UnADB1NF}). In contrast, this behaviour is not observed for BHB-NF or DBHB. In the case of DBHB, the area of the bubble increases with increasing $c$. However, the area of the black hole horizon or disk may decrease or increase (see figures \ref{ABn1},\ref{DAHB1}, and \ref{DADB1}). Enforcing the removal of conical singularities on the bubble and disc (that is, for the BHB-NF and DBHB) seems responsible for this behaviour.  

Finally, Fig \ref{EtaBb2} and Fig \ref{EtaHb} directly compare the distorted the BHB-AF cases by plotting the ratios of the areas of the bubble and horizon respectively at fixed values of the parameters. In Fig \ref{EtaBb2}, the bubble interval length $c-b$ is held fixed on each curve with $a=1$ and we allow $b$ to vary. One sees that the distorted bubble area is initially larger than that of the undistorted case, but as $b$ increases, smoothness forces the distorted bubble to have a smaller area than the undistorted solution. The ratios of horizon areas of the distorted and BHB-AF are plotted in Fig \eqref{EtaHb}. One observes that in both cases, increasing $a$ while holding $c-b = b-a$ fixed has a strong effect on the distorted solution, forcing the areas of the bubble and horizon to increase quickly relative to the undistorted case. Note that $A_1$ and $B_1$ both diverge as $b\rightarrow c$ or $k=1$, so they can no longer be thought of as small.  

\section{Discussion}
An interesting feature of asymptotically flat black hole solutions in spacetime dimensions larger than four is that the domain of outer communication need not be homeomorphic to the exterior of a ball in Euclidean space. As discussed in the Introduction, explicit examples of such solutions have been found in supergravity theories. Such solutions necessarily carry an electric charge. There are no theorems, however, that rule out the existence of pure vacuum solutions. In this work, we have constructed completely regular vacuum solutions in this class by relaxing the asymptotic flatness condition. The solutions contain a non-collapsing $S^2$ `bubble' outside a smooth event horizon with spatial cross-section topology $S^3$ (a completely regular asymptotically flat solution of this type can be constructed in supergravity, although it carries angular momenta and charge \cite{Horowitz:2017fyg}). 

Distorted black holes reveal some unique and remarkable properties, serving as a theoretical framework for understanding how various properties of black holes change when distorted. This paper focuses on examining a 5-dimensional local distorted black hole bubble solution. The metric describing the BHB-AF solution, excluding distortion sources, is presented in a 5-dimensional Weyl form.
The distorted solution is constructed by using the static asymptotically flat Weyl solution as a seed solution. Using the underlying linearity of the vacuum equations reduced on Weyl metrics, we applied `distortions', which can thought of as adding dipole and quadrupole sources to the seed solution. These can be physically interpreted as being produced by distant sources in an asymptotic region far from the event horizon. The distorted DBHB solution is a local solution valid in the region interior to the sources. Distortions influence the black hole horizon. We also have shown that with proper adjustment of the dipole or quadruple distortions we can remove the conical singularities present in the asymptotically flat case.

We expect that constructing explicit asymptotically flat black hole vacuum solutions with non-trivial exterior (i.e. finite spatial rods) will require using inverse scattering techniques that produce non-static solutions with angular momenta. Remarkably, very recently, such a construction was carried for \emph{charged} non-extreme solutions in supergravity \cite{Suzuki:2024phv} to produce charged, asymptotically flat black holes with $L(n,1)$ horizon cross sections. There is strong evidence using integrability methods that regular vacuum solutions might be ruled out \cite{Lucietti:2020ltw}. For example, asymptotically flat vacuum solutions with $L(2,1)$ lens space topology can be effectively ruled out \cite{Lucietti:2020phh}.  Such solutions also have a disc-topology region outside the event horizon, similar to the solutions considered here. It would be interesting to see whether the Weyl distortions method can be applied to a suitable seed solution to produce regular vacuum black lens solutions. 

There are multiple ways to define a black hole, focusing either on the causal structure defining the black hole region to be the set of all points that cannot send signals to future infinity or geometric, defining the black hole region to be the union of all trapped surfaces in a spacetime. At least for dynamical spacetimes, black hole characterization is complicated and these definitions may deviate from each other. It was also shown that highly deformed static distorted Schwarzschild black holes may lack surfaces that are marginally trapped or outer trapping, making the trapping horizon neither future nor outer \cite{TI}. These features have also been analyzed in \cite{B}, where the authors have constructed a black hole deformed by the presence of a thin ring.  Further study of these features of distorted static and stationary black holes could be very insightful.

\paragraph{Acknowledgements}  HKK acknowledges the support of the NSERC Discovery Grant 2018-04887. I.B. acknowledges the support of NSERC Discovery Grant  2018-04873. M.T. was supported by both of
these grants. S.A. acknowledges that this research was supported in part by grant NSF PHY-1748958 to the Kavli Institute for Theoretical Physics (KITP).

\end{document}